\documentclass[utf8]{frontiersinFPHY_FAMS}
\usepackage{url,hyperref,microtype,subcaption,comment}
\usepackage[onehalfspacing]{setspace}
\usepackage{mathrsfs}
\newcommand{\be}{\begin{equation}}
\newcommand{\ee}{\end{equation}}
\newcommand{\bea}{\begin{eqnarray}}
\newcommand{\eea}{\end{eqnarray}}

\newcommand{\MeV}{\text{MeV}}

\newcommand{\mnras}{Mon.~Not.~RAS}

\newcommand{\apj}{ApJ}
\newcommand{\prl}{Phys.~Rev.~Lett.}
\newcommand{\prc}{Phys.~Rev.~C}
\newcommand{\prd}{Phys.~Rev.~D}

\definecolor{red}{rgb}{0.8,0,0}
\definecolor{orange}{rgb}{0.8,0.2,0.0}
\definecolor{blue}{rgb}{0.3,0.0,0.8}
\definecolor{green}{rgb}{0,0.5,0.0}
\definecolor{darkred}{rgb}{0.7,.1,.2}
\definecolor{bgred}{rgb}{1.,.95,.95}
\definecolor{bgblue}{rgb}{.95,.95,1.}
\definecolor{bluegreen}{rgb}{0.,.5,.3}
\definecolor{darkred}{rgb}{0.7,.1,.2}
\definecolor{darkgreen}{rgb}{0.1,.6,.0}
\definecolor{lightyellow}{rgb}{1.,1.,.8}
\definecolor{darkcyan}{rgb}{0.,.7,.9}
\definecolor{lightblue}{rgb}{0.6,0.8,1}
\definecolor{lightgreen}{rgb}{0.7,1.,.9}
\definecolor{money}{rgb}{0.4,0.8,0.}
\definecolor{purple}{rgb}{0.9,0.0,0.8}
\definecolor{orange}{rgb}{0.9,0.5,0.0}
\definecolor{newgr}{rgb}{0.2,0.8,0.2}
\definecolor{newbl}{rgb}{0.3,0.6,0.8}
\definecolor{newor}{rgb}{1.0,0.6,0.}

\usepackage[normalem]{ulem}  

\def\keyFont{\fontsize{8}{11}\helveticabold }
\def\firstAuthorLast{Alford {et~al.}} 
\def\Authors{Mark Alford\,$^{1}$, Arus Harutyunyan\,$^{2,3}$,  Armen Sedrakian\,$^{4,5}$, Stefanos Tsiopelas\,$^{5}$}
\def\Address{$^{1}$Department of Physics, Washington University, St.~Louis,
  Missouri 63130, USA \\
$^{2}$Byurakan Astrophysical Observatory,
Byurakan 0213, Armenia \\
$^{3}$Department of Physics, Yerevan State University, Yerevan 0025,
Armenia \\
$^{4}$Frankfurt Institute for Advanced Studies, D-60438 Frankfurt am Main, Germany\\
$^{5}$Institute of Theoretical Physics, University of Wroc\l{}aw,
50-204 Wroc\l{}aw, Poland 
}
\def\corrAuthor{A. Sedrakian, FIAS, Ruth-Moufang str. 60438, Frankfurt am Main, Germany}
\def\corrEmail{sedrakian@fias.uni-frankfurt.de}
\def\corrAuthor{~}
\def\corrEmail{~}

\begin{document}
\onecolumn
\firstpage{1}

\title[Bulk viscosity and damping timescales]{Bulk Viscosity of Two-Flavor Color Super\-conducting Quark Matter in Neutron Star Mergers} 

\author[\firstAuthorLast ]{\Authors} 
\address{} 
\correspondance{} 

\extraAuth{}

\maketitle

\begin{abstract}
  This work investigates the bulk viscosity of warm, dense,
  neutrino-transparent, color-superconducting quark matter, where
  damping of density oscillations in the kHz frequency range arises
  from weak-interaction-driven direct Urca processes involving
  quarks. We study the two-flavor red-green paired
  color-superconducting (2SC) phase, while allowing for the presence
  of unpaired strange quarks and blue color light quarks of all
  flavors.  Our calculations are based on the SU(3) Nambu–Jona–Lasinio
  (NJL) model, extended to include both vector interactions and the
  't~Hooft determinant term. The primary focus is on how variations in
  the NJL Lagrangian parameters—specifically, the diquark and vector
  coupling strengths—affect both the static properties of quark
  matter, such as its equation of state and composition, and its
  dynamical behavior, including bulk viscosity and associated damping
  timescales. We find that the bulk viscosity and corresponding
  damping timescale can change by more than an order of magnitude upon
  varying the vector coupling by a factor of two at high densities and
  by a lesser degree at lower densities.  This sensitivity primarily
  arises from the susceptibility of 2SC matter, with a smaller
  contribution from modifications to the weak interaction rates. In
  comparison, changes in the diquark coupling have a more limited
  impact. The damping of density oscillations in 2SC matter is similar
  quantitatively to nucleonic matter and can be a leading mechanism of
  dissipation in merging hybrid stars containing color superconducting
  cores.

\tiny
  \keyFont{ \section{Keywords:} Neutron Stars, Neutrino Interactions,
    Quark Matter, Gravitational Waves, Transport
    Coefficients} 
\end{abstract}

\section{Introduction}
The exploration of matter within the extreme environments of neutron
star mergers presents a fascinating way to understand the fundamental
properties of nuclear and quark matter at densities substantially
exceeding the nuclear saturation density. Although recent
gravitational wave observations, such as the GW170817
event~\cite{Abbott2017}, which was accompanied by electromagnetic
counterparts, have only captured the inspiral
 phase of the merger,
 it
is anticipated that the post-merger phase could become observable with
the next generation of gravitational wave detectors.  Nevertheless,
numerical simulations of mergers offer clues into the dynamics of the
merger process and the spectrum of gravitational waves that are
emitted~\cite{Faber2012,Baiotti2017,Baiotti2019,Bauswein2019,Radice2023,Radice2024}.
The gravitational waves, emitted within the first tens to hundreds of
milliseconds after the merger, carry unique information about the
state of matter under extreme conditions. In particular, the post-merger matter is
extremely hot, produces a large number of neutrinos, and, unlike a
supernova, is highly neutron-rich.

The accuracy of the predictions of simulations of binary neutron star
(BNS) mergers depends on a number of factors, among which we would
like to address the need for including dissipation in the typically
nondissipative general-relativistic hydrodynamics.  Dissipation may
control the oscillations that source the gravitational waves. The
possible influence of various transport phenomena has been addressed
in recent years, including the electrical and thermal
conductivities~\cite{Alford2018a,Harutyunyan:2016a,Schmitt:2017efp,Harutyunyan2018b,Harutyunyan2024,Harutyunyan2024PhRvC},
bulk and shear viscosities~\cite{Alford2018a}, etc. In recent years,
much of the attention has been focused on the bulk viscosity-driven
weak $\beta$-decay processes on nucleons~\cite{Alford2019a,
  Alford2020,Alford2021b,Alford2021c,Alford2023},
hyperons~\cite{Alford2021a} and quarks~\cite{Rojas2024, Hernandez2024,
  Alford2024PhRvDLett}.  The microscopic computation of rates allows
one to estimate the damping timescales using local snapshots of matter
conditions provided by
simulations~\cite{Alford2019b,Alford2020,Alford2021c,Alford2023} .

The explicit inclusion of bulk viscosity in numerical simulations is
still an area under active
development~\cite{Most2022,Celora2022,Chabanov2025PhRvL,Chabanov2025PhRvD}. See
also Refs.~\cite{Hammond2021,Radice2022,Camelio2023,Camelio2023-2} for
related studies that provide a more qualitative assessment of bulk
viscous effects based on simulations. Some works, such
as~\cite{Most2022,Celora2022}, account for the bulk viscosity within
frameworks that evolve the system using ideal hydrodynamics, thus
neglecting the back-reaction of the bulk viscosity on the fluid
motion. Other studies incorporate bulk viscosity dynamically, but
assume it to be constant throughout the
evolution~\cite{Chabanov2025PhRvL,Chabanov2025PhRvD}.  Although many
details, such as the dependence of bulk viscosity on temperature,
density, and the composition of matter at supranuclear densities,
remain uncertain, it is generally expected that bulk viscosity can
significantly affect the gravitational wave emission during
inspiral~\cite{Ripley:2023qxo,Ghosh2025} and merger. In particular, it may damp
the oscillations of the remnant stellar core more rapidly, potentially
leading to a faster decay of the gravitational wave signal.

Within this context, the potential presence and behavior of quark
matter remains largely unexplored. Studies of cold quark matter in
hybrid and strange stars, particularly concerning the damping of
$r$-mode oscillations in cold stellar configurations have spanned
several decades~\cite{Madsen1992,Drago2005,Alford:2006gy,Blaschke2007,
  Sad2007a,Sad2007b,Huang2010,Wang2010a,Wang2010b}. Notably,
Ref.~\cite{Alford:2006gy} examined 2SC quark matter at finite
temperatures using a model parameterized by the 2SC gap and chemical
potentials, though without enforcing electric charge neutrality. That
study focused primarily on non-leptonic processes. Importantly,
Ref.~\cite{Alford:2006gy} also derived the bulk viscosity resulting
from coupled non-leptonic and leptonic reactions—an approach similar in spirit to the one used by us
recently~\cite{Alford2024PhRvDLett} and expanded further here.

The investigation of bulk viscosity in quark matter within the context
of BNS mergers remains in its early
stages~\cite{Rojas2024,Hernandez2024,Alford2024PhRvDLett}. Refs.~\cite{Rojas2024,Hernandez2024}
focused on bulk viscosity in unpaired quark matter arising from
non-leptonic and semi-leptonic weak processes. Rojas et
al.\cite{Rojas2024} employed both perturbative QCD and holographic
methods to obtain improved weak- and strong-coupling estimates of the
bulk viscosity. Hernandez et al.~\cite{Hernandez2024} computed the
bulk viscosity using the MIT bag model and perturbative QCD.  Our
previous work~\cite{Alford2024PhRvDLett} presented the first
calculation of bulk viscosity that incorporates the effects of color
superconductivity at intermediate densities and finite temperatures in
the BNS context. In doing so, the conditions specific to BNS mergers,
such as the charge and color neutrality, the density and temperature
dependence of pairing gaps and chemical potentials, were treated
self-consistently within the vector-interaction-enhanced NJL model.
We assume that, in the temperature range $T\le 10$~MeV and for
  densities relevant to the quark cores of neutron stars, the matter
  remains transparent to neutrinos. This assumption does not contradict
  the existing studies of the neutrino mean free path in
  quark matter, although the precise condition for trapping depends on the specific
  model employed—such as the quark matter equation of state (EoS),
  composition, and possible pairing patterns (see
  Refs.~\cite{Carter2000,Steiner2001,Colvero2014}). We
  also expect that the bulk viscosity will be significantly suppressed
  once neutrinos become trapped on microscopic scales and reach
  equilibrium with the surrounding quark matter, consistent with our
  findings in the neutrino-trapped regime of nucleonic
  matter~\cite{Alford2020,Alford2021c,Alford2019b}. 

Refs.~\cite{Rojas2024, Hernandez2024} found a peak in the bulk
viscosity at temperatures $T \leq 0.5$~MeV, which is due to
non-leptonic processes, in agreement with earlier studies of cold
quark matter~\cite{Alford:2006gy}. Furthermore,
Ref.~\cite{Hernandez2024} identified a potential second peak at higher
temperatures, around $T \sim 2$~MeV, which emerges for sufficiently
large strange quark masses and is driven by semi-leptonic
processes. In contrast, Ref.~\cite{Alford2024PhRvDLett} focuses on the
high-temperature regime, $1 \leq T \leq 10$~MeV, and investigates
semi-leptonic processes after verifying that non-leptonic ones
equilibrate too rapidly to contribute significantly at these
temperatures. Their analysis shows a peak within this range, driven by
semi-leptonic Urca processes. Collectively, these studies indicate the
presence of two distinct peaks in the bulk viscosity: the first at low
temperatures ($T \leq 0.5$~MeV) driven by non-leptonic processes, and
the second at higher temperatures (on the order of a few MeV)
associated with semi-leptonic interactions.

This work aims to build upon our previous
study~\cite{Alford2024PhRvDLett}, preserving the essential modeling
framework while focusing on variations of the Lagrangian parameters
within the vector-interaction-enhanced NJL
model~\cite{Bonanno2012}. First, we investigate how the static
composition and EoS of 2SC quark matter with strange
quarks depend on the strengths of the vector and diquark
couplings. Second, we examine dynamical properties, such as the bulk
viscosity and corresponding damping time scales. In this context, we
also analyze how Urca process rates are modified by variations in
pairing strength and the repulsive vector interaction. 

Before proceeding, we point out that the 2SC phase
  studied in this work competes with alternatives, which we briefly
  mention here. One example is the quarkyonic phase which features
  elements of the baryon spectrum for large momenta and quark spectrum
  at small momenta~\cite{McLerran2019,Han2019,Kovensky2020,Kojo2024,Fujimoto2024,Bluhm2025}. The
  quarkyonic phase is expected to occupy the finite-temperature,
  moderate density segment of the QCD phase diagram and its
  competition with the 2SC phase depends on several factors that are
  hard to pin down. Furthermore, it has been realized that the chiral
  phase transition at low temperatures and moderate densities may
  proceed via an inhomogeneous phase, instead of the first-order
  homogeneous transition line~\cite{Buballa2016}. This phase may have
  various realizations such as plane‑wave sinusoidal modulation or
  amplitude‑modulated “kink crystal” condensate, with
  periodically alternating positive and negative values of (real-valued)
  solitonic profile~\cite{Karasawa2016,Abuki2018,Ferrer2022,Tabatabaee2023,Motta2025}.

  Although lattice QCD at finite density is hindered by the sign
  problem, existing results at zero
  density~\cite{Bazavov2012,Aoki2024} allow one to speculate that if
  the $U(1)_A$ anomaly remains strong, it  could suppress the
  emergence of certain exotic phases through modifications of meson
  masses~\cite{Kono2021}. Conversely, if the anomaly becomes weaker at
  finite chemical potential, this may favor the appearance or
  extension of inhomogeneous chiral phases mentioned above.  In
  particular, the axial anomaly contributes to the effective potential
  governing chiral condensate modulations. For spatially inhomogeneous
  phases that rely on chiral spirals or solitonic structures, any
  change in the strength of $U(1)_A$ breaking could alter the relative
  stability of competing condensate
  configurations~\cite{Carignano2020,Gao2022}.  Below we include the
  't Hooft interaction which breaks the $U(1)_A$ symmetry and
  effectively resembles the axial anomaly.  

The present paper is organized as follows. In Sec.~\ref{sec:2SC_phase}
we discuss the equilibrium state of the 2SC phase, including the key
thermodynamic parameters that are required for computing the bulk
viscosity of quark matter. In Sec.~\ref{sec:Urca_rates} we discuss the
computations of the semi-leptonic Urca rates in the 2SC phase.
Sec.~\ref{sec:bulk_visc} presents the general formalism for computing
the bulk viscosity in quark matter based on Urca processes. Our
numerical results are presented in Sec.~\ref{sec:num_results}.  We
conclude and summarize in Sec.~\ref{sec:conclusions}.  Our
calculations use natural units where $\hbar=c=k_B=1$.

\section{Finite temperature 2SC phase}
\label{sec:2SC_phase}

To describe the properties of 2SC quark matter, we adopt a local
vector-interaction-enhanced NJL Lagrangian, which is given by (ignoring electromagnetism) 
\begin{eqnarray}
\label{eq:NJL-Lagrangian}
\mathcal{L}_{NJL}&=&\bar\psi(i\gamma^{\mu}\partial_{\mu}-\hat m)\psi 
+G_S \sum_{a=0}^8 [(\bar\psi\lambda_a\psi)^2+(\bar\psi i\gamma_5 \lambda_a\psi)^2]\nonumber\\
&+& G_V(\bar\psi i \gamma^{\mu}\psi)^2 
+G_D \sum_{\gamma,c}[\bar\psi_{\alpha}^a i \gamma_5
\epsilon^{\alpha\beta\gamma}\epsilon_{abc}(\psi_C)^b_{\beta}][(\bar\psi_C)^r_{\rho} 
i \gamma_5\epsilon^{\rho\sigma\gamma}\epsilon_{rsc}\psi^s_{\sigma}]\nonumber\\
&-&K \left \{ {\rm det}_{f}[\bar\psi(1+\gamma_5)\psi]+{\rm det}_{f}[\bar\psi(1-\gamma_5)\psi]\right\},
\end{eqnarray}
where the quark spinor fields $\psi_{\alpha}^a$ carry color
$a = r, g,b$ and flavor ($\alpha= u, d, s$) indices, the matrix of
quark current masses is given by
$\hat m= {\rm diag}_f(m_u, m_d, m_s)$, $\lambda_a$ where
$ a = 1,..., 8$ are the Gell-Mann matrices in the color space, and
$\lambda_0=(2/3) {\bf 1_f}$, where $ {\bf 1_f}$ is the unit matrix in
the flavor space. The charge conjugate spinors are defined as
$\psi_C=C\bar\psi^T$ and $\bar\psi_C=\psi^T C$, where
$C=i\gamma^2\gamma^0$ is the charge conjugation matrix.  The form of
Eq.~\eqref{eq:NJL-Lagrangian} differs from the original NJL Lagrangian
(which contains only the kinetic energy terms and terms proportional
to $G_S$).  It has been extended by three additional terms: (i) the
vector interaction with coupling $G_V$, which is assumed to be
repulsive, (ii) the pairing interaction with coupling $G_D$, and (iii)
the 't~Hooft interaction with coupling $K$, which breaks the $U_A(1)$
symmetry. The numerical values of the parameters of the Lagrangian are
$m_{u,d}= 5.5$ MeV, $m_s = 140.7$ MeV, $\Lambda = 602.3$ MeV,
$G_S\Lambda^2 =1.835$, $K\Lambda^5 =12.36$~\cite{RehbergPhysRevC}.
These parameters of the SU(3) NJL model were determined by fitting to
empirical data, ensuring that the model accurately reproduces known
properties of mesons and quarks.  The specific physical observables
that are reproduced by the model are the vacuum masses and decay
constants of the pion, kaon, eta-mesons, and the mass of the eta prime
meson, which is particularly sensitive to the $U_A(1)$ anomaly modeled
by the 't~Hooft determinant term.  We note that the temperature range
where the matter is neutrino transparent is limited to $T \lesssim 10$
MeV, which implies that the condition $T/\mu\ll 1$ is always fulfilled
for chemical potentials of quarks of all flavors.  We have
  checked that at the low temperatures considered in this work, there
  are no cutoff artifacts, as the quantities of interest are dominated
  by unpaired quarks near their thermally smeared Fermi surfaces. In
  the temperature regime of interest, the corresponding Fermi
  energies—approximately 300~MeV for light quarks and 500~MeV for
  strange quarks—remain well below the ultraviolet cutoff, placing our
  analysis in a regime where such effects are safely
  suppressed. Nevertheless, should the cutoff approach the Fermi
  surface, a renormalization group–based framework could be employed
  to systematically account for and remove any emerging
  artifacts~\cite{Gholami2025}.  

At intermediate densities, the 2SC phase is expected to be the
dominant pairing channel. More intricate pairing patterns, such as
crystalline and gapless color superconducting states, emerge at the
low temperatures typical of cold neutron stars. However, the 2SC phase
remains a robust feature at finite temperatures. In the 2SC phase,
pairing occurs in a color- and flavor-antisymmetric manner between up
($u$) and down ($d$) quarks, while strange ($s$) quarks remain
unpaired. The pairing gap in the quasiparticle spectrum is
\bea
\Delta_c\propto G_D
\Bigl\langle(\bar\psi_C)_{\alpha}^ai\gamma_5\epsilon^{\alpha\beta
c}\epsilon_{abc}\psi_{\beta}^b\Bigr\rangle\ .
\eea
The overall quark pair wave function must be antisymmetric under
exchange of quarks, according to the Pauli exclusion principle. It is
seen that the color part of the wave function is antisymmetric,
specifically in the anti-triplet configuration $(\bar 3)$ of the SU(3)
color; the flavor part is anti-symmetric as well and involves light up
and down quarks; finally, the spin part is anti-symmetric, implying 
spin-0 pairing, that is, the Cooper pairs form a spin-singlet state.

The quark-antiquark condensates are defined as
\bea
\sigma_\alpha \propto G_S\left\langle\bar{\psi}_\alpha \psi_\alpha\right\rangle,
\eea
and  the constituent mass of each quark flavor is given by
\bea
M_\alpha=m_\alpha-4 G_S \sigma_\alpha+2 K \sigma_\beta \sigma_\gamma.
\eea
The quark dispersion relations include energy shifts arising from the
quark interaction with the vector mean fields $\omega_0$ and $\phi_0$
which are defined as $\omega_0=$
$G_V\langle (\psi_u^{\dagger} \psi_u+\psi_d^{\dagger} \psi_d )\rangle$
and $\phi_0=2 G_V\langle \psi_s^{\dagger} \psi_s\rangle$ and are the
mean field expectation values of the vector mesons $\omega$ and $\phi$
in quark matter.  The quark (thermodynamic) chemical potentials are
given by
\bea
\mu_{f,c}=\frac{1}{3}\mu_B +\mu_Q Q_f+\mu_3T^c_3+\mu_8T^c_8,
\eea
with $\mu_B$ and $\mu_Q$ being the baryon and charge chemical potential, and
\bea
&&Q_f=\operatorname{diag}_f\left(\frac{2}{3},-\frac{1}{3},-\frac{1}{3}\right)
\eea
being the quark charge matrix in flavor space and 
\bea 
&&T_3^c=\frac{1}{2} \operatorname{diag}_c(1,-1,0),\,\, T_8^c=\frac{1}{2 \sqrt{3}} \operatorname{diag}_c(1,1,-2)
\eea
the diagonal generators of the SU(3) color gauge group
related to the 3rd and 8th gluons. 
The values of $\mu_Q$, $\mu_3$, and $\mu_8$ are determined by the requirement of electrical and color neutrality.
For some purposes, such as calculating Fermi-Dirac factors, these
can be absorbed into ``effective'' quark chemical potentials
\bea\label{eq:chem_pot_eff}
{\mu}^*=\operatorname{diag}_f\left(\mu_u-\omega_0, \mu_d-\omega_0,
  \mu_s-\phi_0\right).
\eea
Starting from the Lagrangian \eqref{eq:NJL-Lagrangian} the partition
function and the thermodynamic potential $\Omega$ of the 2SC phase can
be computed in the mean-field approximation; see, for example,
Refs.~\cite{Alford2008RvMP,Ruster2005PhRvD,Blaschke2005PhRvD,Gomez2006PhRvD,Bonanno2012}. We
find the values of the chemical potentials $\mu_Q$, $\mu_3$, $\mu_8$
by requiring neutrality, i.e. setting
$\partial\Omega/\partial\mu_i=0$. The mean fields, including the
pairing gaps, are also obtained by stationarizing $\Omega$ with
respect to their values. For the computation of the incompressibility
of quark matter, we will need the expression for the pressure which is
given by
\bea\label{eq:Pressure}
 P  &=& \frac{1}{2 \pi^2} \sum_{i=1}^{18} \int_0^{\Lambda} d k k^2
\left[\left|\epsilon_i\right|+2 T \ln \left(1+\mathrm{e}^{-\frac{\left|\epsilon_i\right|}{T}}\right)\right]
 +4 K \sigma_u \sigma_d \sigma_s 
 -\frac{1}{4 G_D} \sum_{c=1}^3\left|\Delta_c\right|^2
 \nonumber\\
 &-&2 G_S \sum_{\alpha=1}^3 \sigma_\alpha^2+\frac{1}{4 G_V}\left(2 \omega_0^2+\phi_0^2\right) +
 \sum_{l=e^{-}, \mu^{-}} P_l-P_0-B^*,
\eea
where $\epsilon_i$ are the quasiparticle energies of quarks in quark
matter, $P_l$ is lepton pressure (which is approximated as
corresponding to an ideal relativistic gas), $P_0$ is the vacuum
pressure, and $B^*$ is an effective bag constant, which controls to
deconfinement phase transition density.

In the following, we will explore how the Urca rates and resultant
bulk viscosity depend on key parameters in the Lagrangian. In doing
so, we will vary the temperature in the range $1\le T\le 10$~MeV, in
which we assume the neutrinos are free-streaming. As a main point of
this work, we will vary the two additional couplings in the Lagrangian
that enhance the ordinary NJL model. Specifically, to understand the
role of pairing interaction strength, we consider two values of
diquark coupling $G_D/G_S = 1$ and 1.25.  Similarly, we vary the
vector interaction in the range $0.6 \le G_V/G_S \le 1.2$ to highlight
its role in the physics of the bulk viscosity of 2SC matter.
\begin{figure}[bt]
\begin{center}
\includegraphics[width=17cm]{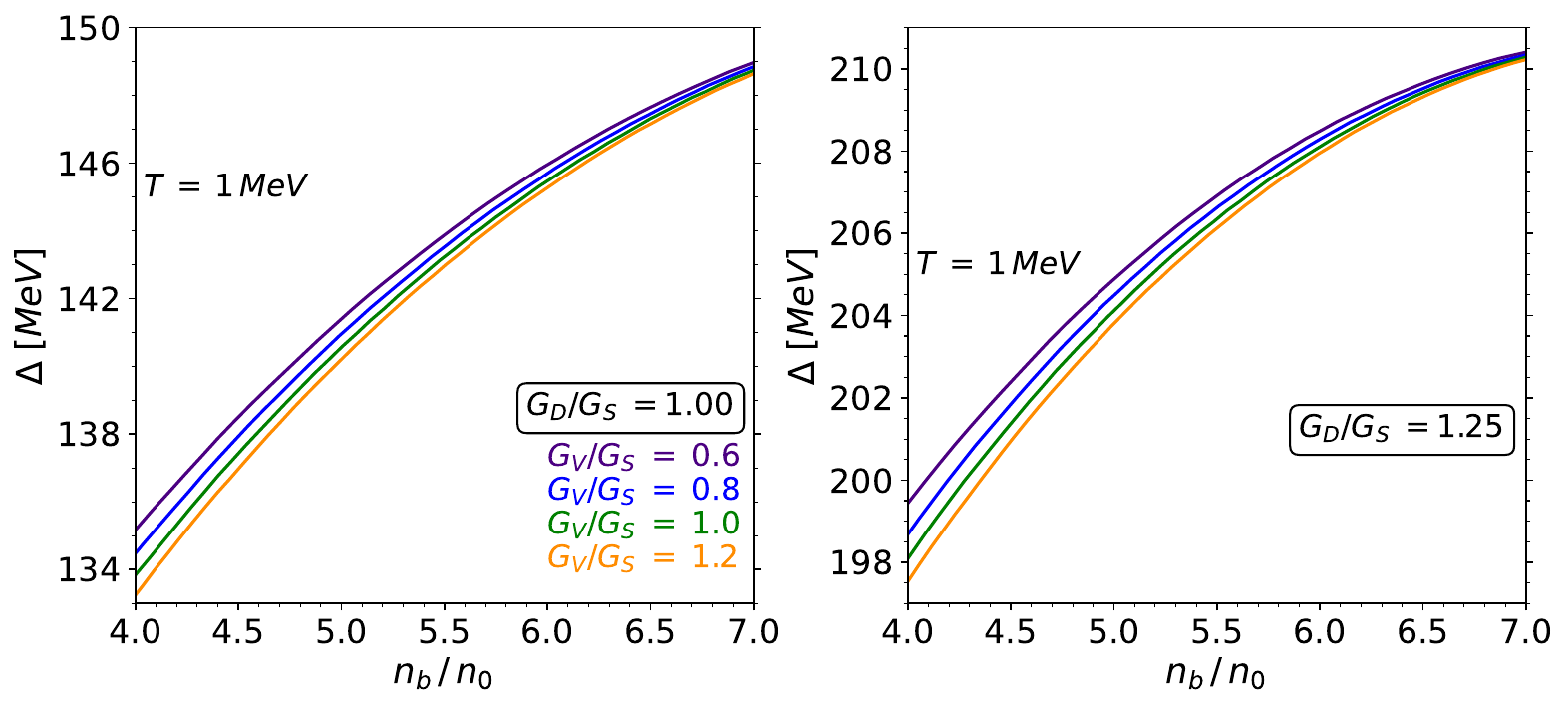}
\end{center}
\caption{2SC gap as a function of number density for temperature $T=1$
  MeV and varying values of vector and diquark couplings. The
  temperature dependence of the gap is weak in the regime of interest,
  where $T\ll \Delta$.  }
\label{fig:gaps}
\end{figure}
\begin{figure}[h!]
\begin{center}
\includegraphics[width=17cm]{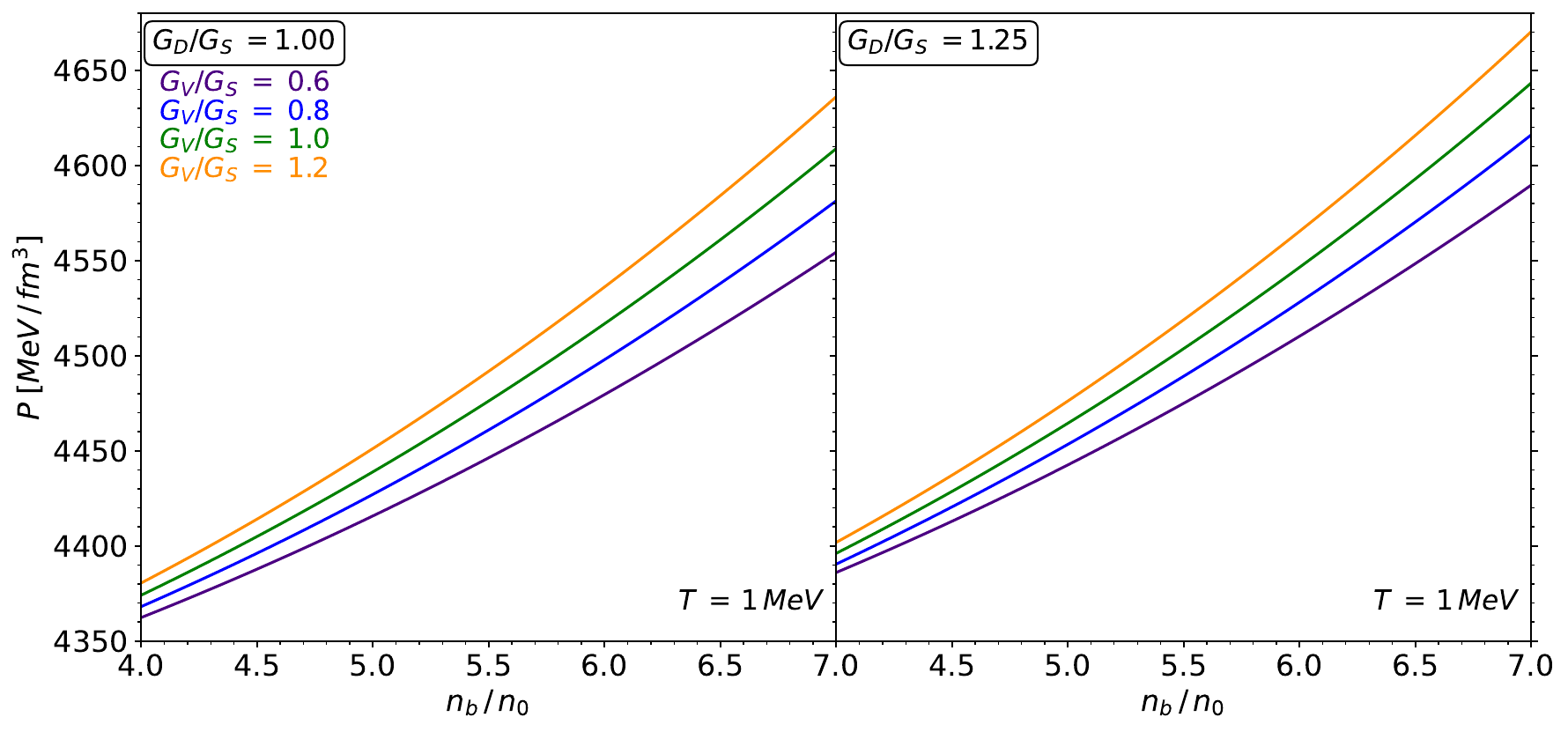}
\end{center}
\caption{Pressure as a function of number density for
  temperature $T=1$ MeV and various fixed values of vector and diquark couplings.}
\label{fig:pressure}
\end{figure}

Figure~\ref{fig:gaps} shows the density dependence of the gap for
fixed temperature $T=1$~MeV and several values of $G_D/G_S$ and
$G_V/G_S$ in the range indicated above.  The gaps predicted by the
model are much larger than the characteristic temperatures relevant
for untrapped neutrinos, i.e., we are working essentially in the limit
$T\ll \Delta$. Therefore, the excitations are exponentially
suppressed, and the variations of the gap with temperature are
insignificant.  The enhancement of the gap with increasing density may
be associated with the increase of the density of states at the Fermi
surfaces of quarks, while coupling being fixed.  Further, it is seen
that the increase of the attractive pairing strength $G_D/G_S$ from 1
to 1.25 increases the gap by $\ge 10\%$. Finally, the repulsive vector
interaction acts oppositely by reducing the pairing gap by at most a
few percent when going from $G_V/G_S= 0.6$ to $G_V/G_S=1.2$.  The
energy gap indicates how effectively the phase space of green and red
light quarks is suppressed in weak reactions, thereby limiting their
contribution to bulk viscosity. The gap variations across model
parameters seen in Fig.~\ref{fig:gaps} consistently maintain the
effective shutdown of these quark color-flavor channels.

Figure~\ref{fig:pressure} shows the pressure of the NJL model for
fixed temperature $T=1$~MeV and several values of $G_D/G_S$ and
$G_V/G_S$.  As expected, the pressure increases as both the diquark
and vector couplings are increased. The significance of the pressure
in the present context is that its derivative with respect to baryon
density enters the computation of compressibility of matter, which
enters the expression for damping timescale via the bulk viscosity,
see Eq.~\eqref{eq:damping_time} below.  Note that we use in
Eq.~\eqref{eq:Pressure} the value $B^*=0$ as we are interested only in
the pressure derivative.

\begin{figure}[hbt]
\begin{center}
\includegraphics[width=17cm]{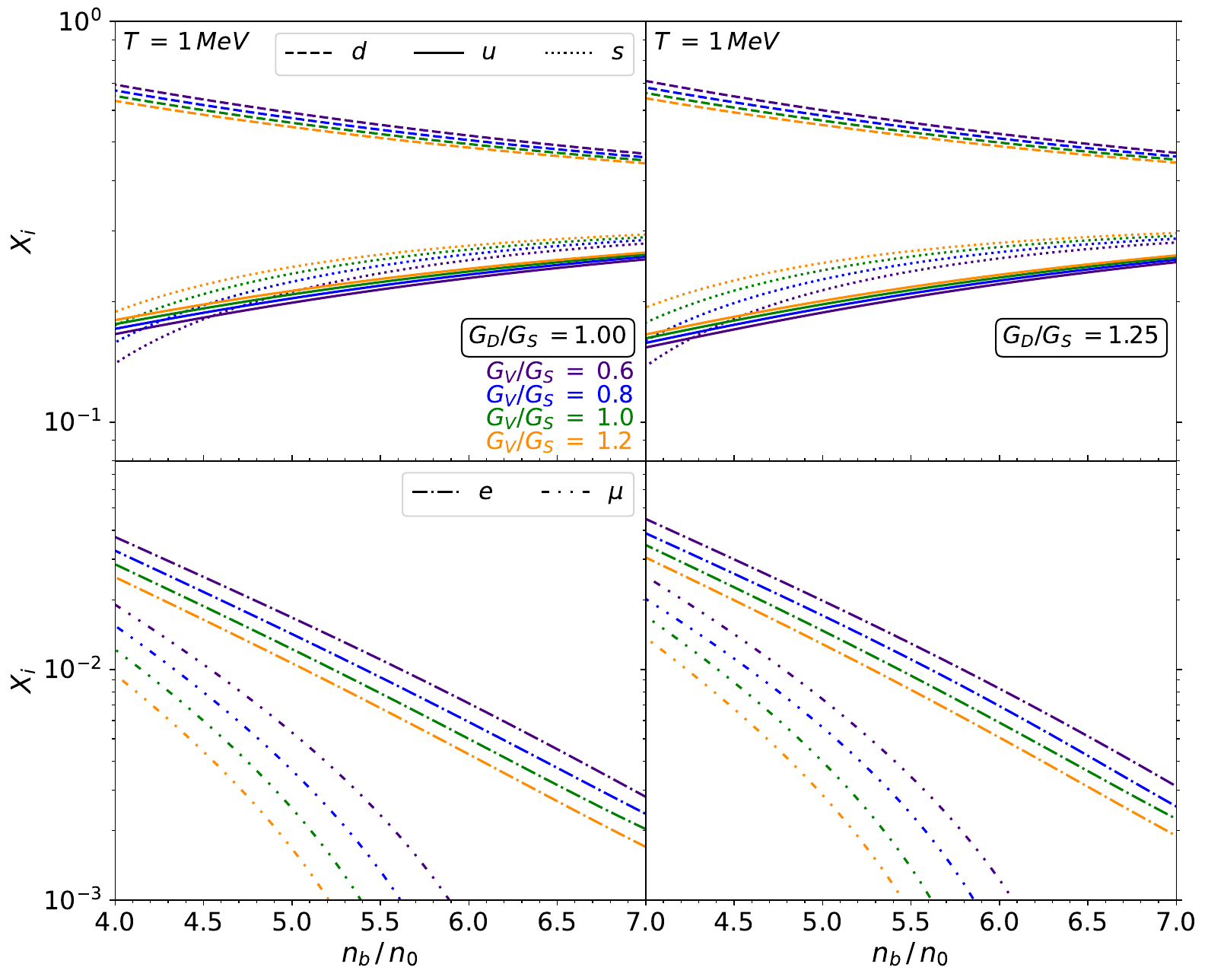}
\end{center}
\caption{Composition of 2SC matter as a function of number density for
  $T=1$~MeV and varying vector and diquark couplings, where $X_i =
  n_i/n_b$ with $n_i$ being the density of any given species. Top row:
  unpaired (blue) quark fractions. Bottom row: charged lepton ($e^-$
  and $\mu^-$) fractions.}\label{fig:composition}
\end{figure}

Figure~\ref{fig:composition} displays the composition of the 2SC phase
in $\beta$-equilibrium for varying diquark and vector couplings.  It
focuses on the populations of unpaired (blue-colored) up and down
quarks, along with blue strange quarks, which are relevant for the
computation of the bulk viscosity.  Note that blue strange quarks have
slightly different densities compared to their red-green counterparts,
with chemical potential differences of approximately $2\%$ (roughly 10
MeV).  The substantial strange quark population rapidly equilibrates
with the light flavor sector through non-leptonic processes.
Additionally, strange quarks are shown to reduce the lepton fraction
in matter, a phenomenon that was identified decades
ago~\cite{Duncan1983ApJ,Duncan1984ApJ}.  Examining the variations in
coupling constants, we observe that increasing the attractive pairing
strength $G_D/G_S$ from 1 to 1.25 slightly raises the $d$-quark
population while reducing the $u$-quark population, accompanied by
increased electron and muon populations. Additionally, strengthening
the repulsive vector interaction increases the strange and $u$-quark
populations while reducing the $d$-blue quark populations. Through
$\beta$-equilibrium, electron and muon populations are suppressed as
the vector coupling increases.

\begin{figure}[bt]
\begin{center}
\includegraphics[width=17cm]{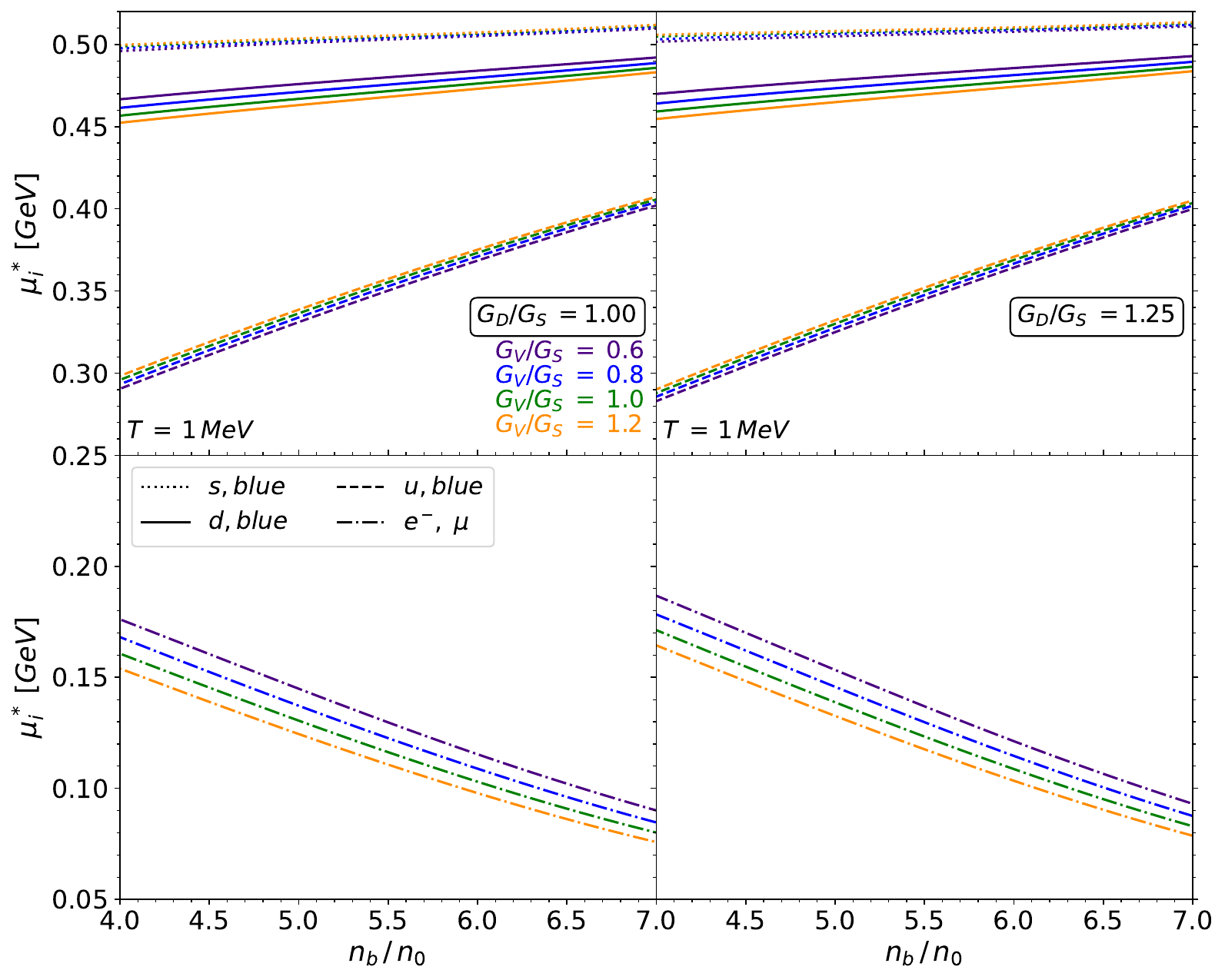}
\end{center}
\caption{Effective chemical potentials as functions of number density
  for $T=1$ MeV and various fixed values of the vector and diquark
  couplings. The variations of chemical potentials with temperature
  are insignificant in the range of interest and are not shown.  }
\label{fig:chem_pot}
\end{figure}

\begin{figure}[h!]
\begin{center}
\includegraphics[width=17cm]{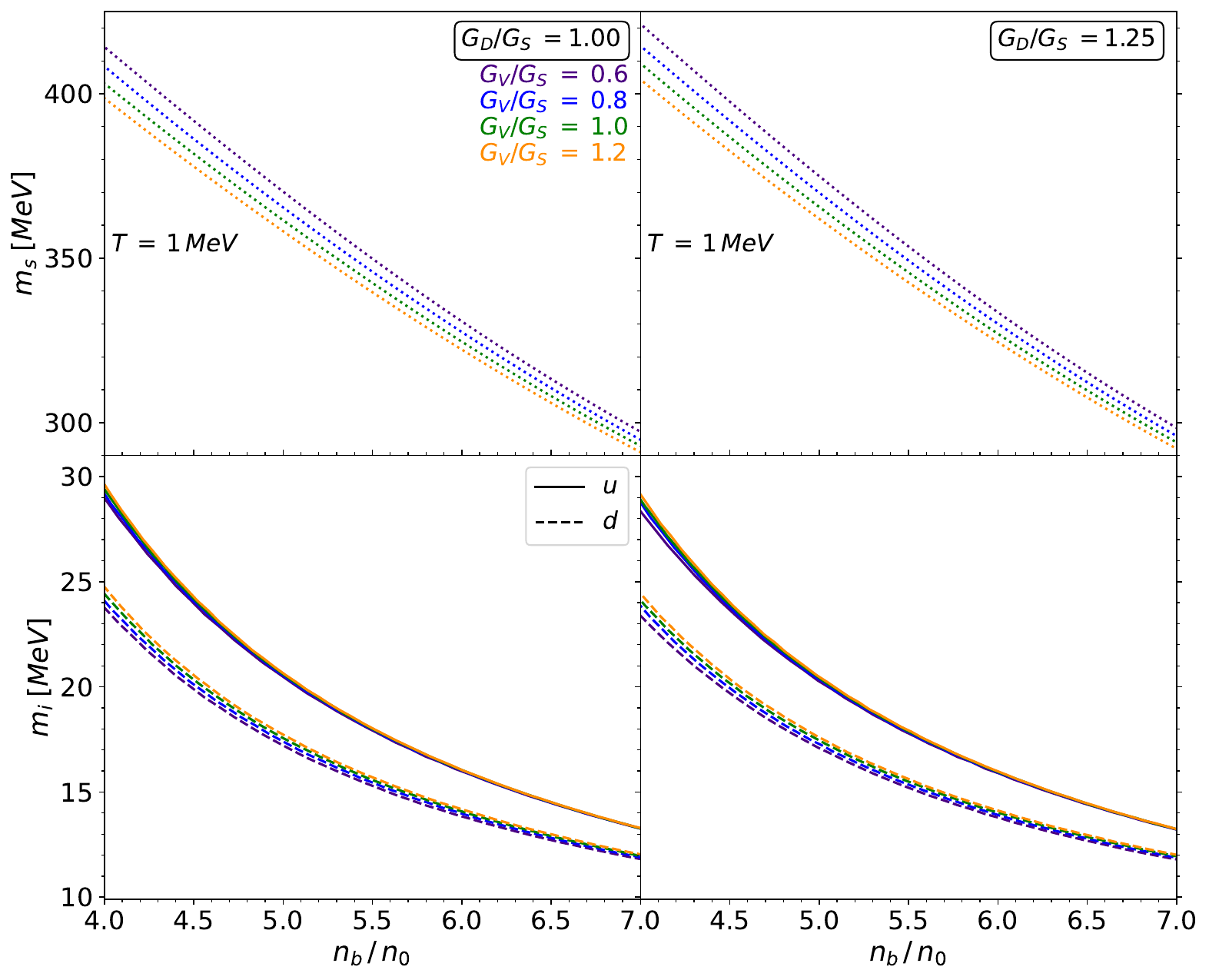}
\end{center}
\caption{Quark masses as functions of number density for $T=1$~MeV and
  various fixed values of vector and diquark couplings. The variations
  of masses with temperature are insignificant in the range of
  interest and are not shown.
\label{fig:masses} }
\end{figure}

Before discussing the reaction rates, we first examine the (effective)
chemical potentials and masses of quarks and leptons shown in
Figs.~\ref{fig:chem_pot} and \ref{fig:masses}. For quarks, the masses
are modified by medium effects through the chiral condensate, which
generally depends on both density and temperature. This comparison
reveals to what degree the different particles are
relativistic. Nevertheless, the formalism applied in the following
sections is fully relativistic.  Figure~\ref{fig:chem_pot} shows the
effective chemical potentials defined by \eqref{eq:chem_pot_eff} as
functions of density for fixed values of temperatures as well as
coupling constants $G_D/G_S$ and $G_V/G_S$.  We show only the chemical
potentials of blue quarks, which are relevant for the computation of
the bulk viscosity. The temperature dependence of the chemical
potentials is weak in the shown density range. In addition, the lower
panels show the chemical potentials of electrons and muons, which are
connected through $\beta$-equilibrium conditions to the chemical
potentials of quarks. The general trend is that the quark chemical
potentials rise with density, as one would expect, and the lepton
chemical potentials drop, as the rising fraction of strange quarks
means that fewer charged leptons are needed to establish electrical
neutrality.  Rapid non-leptonic processes enforce down-strange flavor
equilibrium, ensuring $\mu_d=\mu_s$. Therefore, the observed
differences between the effective chemical potentials of $d$ and $s$
quarks in Fig.~\ref{fig:chem_pot} arise from their respective
couplings to $\omega$ and $\phi$ mesons.  Increasing the pairing
interaction increases the chemical potentials of strange and down
quarks while reducing that of up quarks. Simultaneously, the chemical
potentials of electrons and muons (which are equal) also increase.
Finally, increasing the repulsive vector interaction increases
$s$-quark and $u$-quark chemical potentials and decreases the
$d$-quark chemical potential. In parallel, through
$\beta$-equilibrium, the electron and muon chemical potentials
decrease.

Fig.~\ref{fig:masses} shows the masses of blue quarks as functions of
density.  An increase in the diquark coupling leads to a rise in the
mass of the $s-$ and a decrease in the dynamically generated masses of
$d$- and $u$-quarks. Additionally, an increase in the vector coupling
reduces the $s$-quark mass and raises the $u$- and $d$-quark
masses. The temperature dependence of the quark masses is observed to
be very weak. Comparing effective quark chemical potentials shown in
Fig.~\ref{fig:chem_pot} to the respective masses of quarks, we
conclude that the light quarks are ultrarelativistic in the entire
density range considered. The strange quarks are mildly relativistic
in the low-density limit $n_b\simeq n_0$, where $n_0$ is the
saturation density, but become strongly relativistic as the density
increases.

\section{Urca reaction rates for $ude$  and $udse$ compositions}
\label{sec:Urca_rates} 

We examine neutrino-transparent $ude$ and $udse$ matter, i.e.,
  matter consisting of $u, d$ or $u,d, s$ quarks and electrons. Muons
  are also included in the composition whenever they become
  energetically favorable.  Muonic contribution to the reaction rates
is subdominant and will be neglected below; the case of nuclear matter
has been studied in Ref.~\cite{Alford2023}.  The fundamental
(semi-leptonic) $\beta$-equilibration processes in this system include
$d$ and $s$-quark decay along with the electron capture processes of
the direct Urca type:
\bea\label{eq:d_decay}
&& d\rightarrow u + e^-+\bar{\nu}_e,\\
\label{eq:e_capture_d}
&& u + e^-\rightarrow d +{\nu}_e,\\
\label{eq:s_decay}
&& s\rightarrow u + e^-+\bar{\nu}_e,\\
\label{eq:e_capture_s}
&& u + e^-\rightarrow s +{\nu}_e,
\eea
where $\nu_e$ and $\bar\nu_e$ are the electron neutrino and
antineutrino, respectively. The
processes~\eqref{eq:d_decay}--\eqref{eq:e_capture_s} proceed
exclusively from left to right because in neutrino-transparent matter,
neutrinos/antineutrinos can only appear in final states. The rates of
the Urca processes \eqref{eq:d_decay}--\eqref{eq:e_capture_s} can be
written as
\bea \label{eq:Gamma1_def}
\Gamma_{d/s\to ue\bar\nu} &=& \int\!\! \frac{d^3p}{(2\pi)^32p_0} \int\!\!
\frac{d^3p'}{(2\pi)^32p'_0} \int\!\! \frac{d^3k}{(2\pi)^32k_0}
\int\!\! \frac{d^3k'}{(2\pi)^32k'_0}\sum \vert 
{\mathscr M}_{\rm Urca}\vert^2 \nonumber\\
& \times & \bar{f}(k)\bar{f}(p)
f(p') (2\pi)^4\delta^{(4)}(k+p+k'-p'),\\
\label{eq:Gamma2_def}
\Gamma_{ue\to (d/s)\nu} &=& \int\!\! \frac{d^3p}{(2\pi)^32p_0} \int\!\!
\frac{d^3p'}{(2\pi)^32p'_0} \int\!\! \frac{d^3k}{(2\pi)^32k_0}
\int\!\! \frac{d^3k'}{(2\pi)^32k'_0}
\sum \vert {\mathscr M}_{\rm Urca}\vert^2 \nonumber\\
& \times & {f}(k) {f}(p) 
\bar{f}(p')
(2\pi)^4\delta(k+p-k'-p').
\eea
where $f(p) = \{\exp[(E_p-\mu^*)/T+1\}^{-1}$ etc. are the Fermi
distribution functions of fermions (quarks and leptons) with
$E_p=\sqrt{p^2+m^{2}}$ being the single-particle spectrum of particles
with mass $m$, and $\bar{f}(p)=1-f(p)$. The effective chemical
potentials for quarks are given by Eq.~\eqref{eq:chem_pot_eff}, and
for leptons we have simply $\mu^*_l=\mu_l$.  The mapping between the
particle labeling and their momenta is as follows: $(e) \to k$,
$(\nu/\bar{\nu}) \to k'$, $(u) \to p$, and $(d/s) \to p'$.

The spin-averaged relativistic matrix element of the Urca processes reads 
\be\label{eq:matrix_el_full}
\sum \vert {\mathscr M}_{\rm Urca}\vert^2 = 128 G_F^2\cos^2
\theta_c (k\cdot p) (k'\cdot p'),
\ee 
where $G_F=1.166\cdot 10 ^{-5}$ GeV$^{-2}$ is the Fermi coupling
constant and $\theta_c$ is the Cabibbo angle with
$\cos\theta_c=0.974$.  For the matrix element of the Urca processes
including $s$-quark, one simply needs to replace $\cos\theta_c$ with
$\sin\theta_c$.  The twelve-dimensional phase-space integrals in
Eqs.~\eqref{eq:Gamma1_def} and \eqref{eq:Gamma2_def} can then be
reduced to four-dimensional integrals. We follow the method of
Ref.~\cite{Alford2021c}.

Before doing so, we note that, as has previously been noted
\cite{Alford:2006gy,Rojas2024,Hernandez2024}, the non-leptonic
processes
\bea\label{eq:non-leptonic}
u+d\leftrightarrow u+s
\eea
equilibrate much faster than the other Urca processes.  As a result,
in the regime where the semi-leptonic Urca process contributes to the
bulk viscosity significantly, the non-leptonic equilibration is
already completed and one can assume $\mu_s=\mu_d$ at all times
(see~\cite{Alford:2006gy}).  Under this condition, we have
only a single equilibrating quantity
$\mu_\Delta\equiv\mu_d-\mu_u-\mu_e$,
which is the relevant measure of how much the system is driven out of
$\beta$-equilibrium state by a cycle of compression and rarefaction.

Returning to Eqs.~\eqref{eq:Gamma1_def} and \eqref{eq:Gamma2_def},
we carry out part of the integrations, as described in Ref.~\cite{Alford2021c}, and obtain
\bea\label{eq:Gamma1_final} 
\Gamma_{d/s\to ue\bar\nu} (\mu_{\Delta})
&=& -\frac{{G}^2T^4}{8\pi^5} 
\int_{-\infty}^\infty\!\!\! dy\,
\!\int_0^\infty\!\! dx\, \left[(\mu_{d/s}^*+yT)^2 
 -m_{d/s}^{2}-x^2T^2\right]\nonumber\\
 &&\hspace{-3cm}
\left[(\mu_l +\mu_u^* +\bar{y}T)^2
-m_e^2-m_u^{2} -x^2T^2\right]
\int_{m_e/T-\alpha_e}^{\alpha_u+\bar{y}}\! 
dz\, \bar{f}(z){f}(z-\bar{y})\,\theta_x\! 
\int_{0}^\infty\! dz'\,f(z'+y)\,\theta_y,\\
\label{eq:Gamma2_final} 
\Gamma_{ue\to (d/s)\nu} (\mu_{\Delta}) 
&=& \frac{{G}^2T^4}{8\pi^5}\int_{-\infty}^\infty\!
dy\! \int_0^\infty\! dx\, \left[(\mu_{d/s}^*+yT)^2 
  -m_{d/s}^{2}-x^2T^2\right]\nonumber\\
&&\hspace{-3.5cm}\left[(\mu_e +\mu_u^* +\bar{y}T)^2
-m_e^2-m_u^{2} -x^2T^2\right]
\int_{m_e/T-\alpha_e}^{\alpha_u +\bar{y}}\! 
dz\, f(z)f(\bar{y}-z)\,\theta_x\! \int_{0}^{\alpha_d+y}\! 
dz'\, {f}(z'-y)\,\theta_z,
\label{eq:ud22}
\eea
where $G=G_F \cos\theta_c$, $m_e$ is the electron mass,
$\alpha_i = \mu^*_i/T=\mu_i/T-u_i$, with $u_d=u_u=\omega_0/T$, $u_s=\phi_0/T$ and $u_e=u_\nu=0$;
$\bar{y}=y+\mu_{\Delta}/T$ with $\mu_{\Delta}=\mu_d-\mu_u-\mu_e$, and $f(x) = (e^x+1)^{-1}$ is the Fermi distribution
function of the dimensionless variable $x$.
The $\theta$-functions in Eqs.~\eqref{eq:Gamma1_final} and 
\eqref{eq:Gamma2_final} imply 
\bea\label{eq:thetax}
\theta_x &: &
(z_k-x)^2 \leq \left(z -\alpha_u
-\bar{y}\right)^2 -m_u^{2}/T^2\leq (z_k+x)^2,\\
\label{eq:thetay}
\theta_y &: &
(z'-x)^2 \leq \left(z' +\alpha_{d/s}+ y\right)^2 
-m_{d/s}^{2}/T^2\leq (z'+x)^2,\\
\label{eq:thetaz}
\theta_z &: &
(z'-x)^2 \leq \left(z'-\alpha_{d/s}-y\right)^2 
-m_{d/s}^{2}/T^2\leq (z'+x)^2.
\eea 
The integration variables $y$ and $x$ are normalized-by-temperature
transferred energy and momentum, respectively; the variable $z$ is the
normalized-by-temperature electron energy, computed from its chemical
potential, $z_k=\sqrt{(z+\alpha_l)^2 -m_l^2/T^2}$ is the normalized
lepton momentum, and $z'$ is the normalized neutrino/antineutrino
energy.

Alternative forms of Eqs.~\eqref{eq:Gamma1_final} and
\eqref{eq:Gamma2_final} can be obtained by exploiting the relation
$f(x)f(y)=g(x+y)[1-f(x)-f(y)]$ after which the inner two integrals can
be done analytically. We find then
\bea\label{eq:Gamma_ddecay}
\Gamma_{d/s\to u e \bar\nu} (\mu_{\Delta}) 
&=& 
\frac{{G}^2T^8}{8\pi^5} 
\int_{-\infty}^\infty\!\!\! dy\, 
[1+g(\bar{y})] \!\int_0^\infty\!\! dx \left[(\alpha_e
  +\alpha_u+\bar{y})^2-x^2-(m_e^2+m_u^{2})/T^2\right]
\nonumber\\
&&\hspace{-3.7cm}
\left[(\alpha_{d/s}+y)^2 
-x^2-m_{d/s}^{2}/T^2\right]
\ln \Bigg\vert \frac{1+\exp\left(-z_{2\rm max}-y\right)}
{1+\exp\left(-z_{2\rm min}-y\right)}\Bigg\vert
\ln \Bigg\vert \frac{1+\exp\left(-z_{1\rm min}\right)} {1+\exp\left(-z_{1\rm min}+\bar{y}\right)}
\frac{1+\exp\left(-z_{1\rm max}+\bar{y}\right)} {1+\exp\left(-z_{1\rm
max}\right)}\Bigg\vert,
\nonumber\\\\
\label{eq:Gamma_ecapture}
\Gamma_{ue\to (d/s)\nu} (\mu_{\Delta}) 
&=& 
\frac{{G}^2T^8}{8\pi^5}\int_{-\infty}^\infty\!
dy\, g(\bar{y}) \! \int_0^\infty\! 
dx \left[(\alpha_e +\alpha_u
  +\bar{y})^2-x^2-(m_e^2+m_u^{2})/T^2\right]
\nonumber\\
&&\hspace{-3.7cm}\left[(\alpha_{d/s}+y)^2 
-x^2-m_{d/s}^{2}/T^2\right]
\ln \Bigg\vert \frac{1+\exp\left(-z_{3\rm max}+y\right)}
{1+\exp\left(-z_{3\rm min}+y\right)}\Bigg\vert 
\ln \Bigg\vert \frac{1+\exp\left(-z_{1\rm min}\right)} {1+\exp\left(-z_{1\rm min}+\bar{y}\right)}
\frac{1+\exp\left(-z_{1\rm max}+\bar{y}\right)} {1+\exp\left(-z_{1\rm max}\right)}\Bigg\vert.
\nonumber\\
\eea
The energy integration limits $z_{i\max}$ and $z_{i\min}$ are
determined by solving Eqs.~\eqref{eq:thetax}, \eqref{eq:thetay} and
\eqref{eq:thetaz} and then comparing the results to the initial energy
integration bounds specified in Eqs.~\eqref{eq:Gamma1_final} and
\eqref{eq:Gamma2_final}.

As discussed above, the light-flavor quarks are ultrarelativistic
under the considered conditions, therefore, we take this limit in
Eqs.~\eqref{eq:thetax}, \eqref{eq:thetay} and \eqref{eq:thetaz}, after
which they simplify to (note that $z, z', y\sim 1\ll x, \alpha_i$)
\bea\label{eq:thetax1}
\theta_x &: &
-z-\alpha_e+x \leq -z +\alpha_u
+\bar{y} \leq z+\alpha_e+x\quad \Rightarrow \quad
-2\alpha_e\leq -\alpha_e+\alpha_u
+\bar{y} -x\leq 2z,\quad\\
\label{eq:thetay1}
\theta_y &: &
-z'+x \leq z' +\alpha_{d}+ y \leq z'+x\quad \Rightarrow \quad 0\leq x-\alpha_{d}-y\leq 2z',\\
\label{eq:thetaz1}
\theta_z &: &
-z'+x \leq -z'+\alpha_{d}+y \leq z'+x\quad \Rightarrow \quad 0\leq \alpha_{d}+y -x\leq 2z',
\eea 
which, together with the limits of integration in
Eqs.~\eqref{eq:Gamma1_final} and \eqref{eq:Gamma2_final}, imply
\bea\label{eq:thetax4}
\theta_x &: &
\theta(\alpha_e+\alpha_u
+\bar{y} -x),
\quad z_{1\rm min}=\frac{-\alpha_e+\alpha_u+\bar{y} -x}{2},\quad
z_{1\rm max}=\alpha_u+\bar{y},\\
\label{eq:thetay4}
\theta_y &: &
\theta(x-\alpha_d-y),\qquad
z_{2\rm min}=\frac{x-\alpha_d-y}{2},\quad
z_{2\rm max}=\infty,\\
\label{eq:thetaz4}
\theta_z &: &
\theta(\alpha_d+y -x),\qquad
z_{3\rm min}=\frac{\alpha_d+y -x}{2},\quad
z_{3\rm max}=\alpha_d+y.
\eea 
Further simplifications of Eqs.~\eqref{eq:Gamma_ddecay} and
\eqref{eq:Gamma_ecapture} can be achieved in the low temperature
limit, which corresponds to $\alpha_i\gg 1$. In this case
$z_{1\rm min}\to -\infty$, $z_{1,2,3\rm max}\to +\infty$, but
$z_{2,3\rm min}\sim y\sim 1 $ because $x\simeq \alpha_d$. Using the
limiting formula
\bea\label{eq:log_limit}
\lim_{\vert z\vert \to\infty }L(z,y)\equiv
\lim_{\vert z\vert \to\infty }\ln \Bigg\vert
\frac{1+\exp\left(-z\right)} {1+\exp\left(-z-y\right)}\Bigg\vert =  y\theta (-z),
\eea 
we can approximate the logarithms in Eqs.~\eqref{eq:Gamma_ddecay} and
\eqref{eq:Gamma_ecapture} as 
\bea\label{eq:log1_limit}
\lim_{\vert z\vert \to\infty }L_1 &= &\lim_{\vert z\vert \to\infty }[ L(z_{1\rm min},-\bar{y})-
L(z_{1\rm max},-\bar{y})]
\simeq 
-\bar{y}\theta (-z_{1\rm min}) = -\bar{y},\\
\label{eq:log2_limit}
\lim_{\vert z\vert \to\infty }L_2 &=& 
\lim_{\vert z\vert \to\infty }\ln \Bigg\vert
\frac{1+\exp\left(-z_{2\rm max}-y\right)} {1+\exp\left(-z_{2\rm min}-y\right)}\Bigg\vert
\simeq  -\ln \vert 1+\exp\left(-z_{2\rm min}-y\right)\vert,\\
\label{eq:log3_limit}
\lim_{\vert z\vert \to\infty }L_3 &=& 
\lim_{\vert z\vert \to\infty }\ln \Bigg\vert
\frac{1+\exp\left(-z_{3\rm max}+y\right)} {1+\exp\left(-z_{3\rm min}+y\right)}\Bigg\vert
\simeq -\ln \vert 1+\exp\left(-z_{3\rm min}+y\right)\vert,
\eea 
after which we obtain
\bea\label{eq:Gamma_ddecay1}
\Gamma_{d\to u e \bar\nu} (\mu_{\Delta}) 
&=& 
\frac{{G}^2T^8}{8\pi^5} 
\int_{-\infty}^\infty\!\!\! dy\, 
\bar{y}[1+g(\bar{y})] \!\int_{\alpha_d+y}^{\alpha_e+\alpha_u
  +\bar{y}}\! dx \left[(\alpha_e +\alpha_u+\bar{y})^2-x^2\right] \nonumber\\
&&\times \left[(\alpha_d+y)^2 -x^2\right]
\ln \bigg\vert 1+\exp\left(-\frac{x+y-\alpha_d}{2}\right)\bigg\vert 
\theta (\alpha_e+\alpha_u+\bar{y}-\alpha_d-y),\quad\\
\label{eq:Gamma_ecapture1}
\Gamma_{ue\to d\nu} (\mu_{\Delta}) 
&=& 
\frac{{G}^2T^8}{8\pi^5}\int_{-\infty}^\infty\!
dy\,\bar{y} g(\bar{y}) \! \int_0^{{\rm min}\{\alpha_d+y;\alpha_e+\alpha_u+\bar{y}\}}\!
dx \left[(\alpha_e +\alpha_u +\bar{y})^2-x^2\right]\nonumber\\
&&\times \left[(\alpha_d+y)^2 -x^2\right]
\ln \bigg\vert 1+\exp\left(\frac{x+y-\alpha_d}{2}\right)\bigg\vert.
\eea
It is interesting to note that these expressions can be rewritten in
terms of the shift between the chemical potentials of $u$- and
$d$-quarks $\Delta u=u_d-u_u$ (this shift vanishes in the case
discussed here as $u$ and $d$ quarks are coupled to the $\omega$-meson
due to Eq.~\eqref{eq:chem_pot_eff}, but it might be nonzero if
isovector channel of interaction is included in analogy to the
$\rho$-meson in the case of nucleonic matter discussed in
Ref.~\cite{Alford2021c}). Indeed, using the relation
\bea
\alpha_e+\alpha_u+\bar{y}=\alpha_d+y+\Delta u, 
 \eea
we obtain 
\bea\label{eq:Gamma_ddecay2}
\Gamma_{d\to u e \bar\nu} (\mu_{\Delta}) 
&=& 
\frac{{G}^2T^8}{8\pi^5} \theta (\Delta u)
\int_{-\infty}^\infty\!\!\! dy\, 
\bar{y}[1+g(\bar{y})] \!\int_{\alpha_d+y}^{\alpha_d +y +\Delta u}
\! dx \left[(\alpha_d +y +\Delta u)^2-x^2\right] \nonumber\\
&&\times \left[(\alpha_d+y)^2 -x^2\right]
\ln \bigg\vert 1+\exp\left(-\frac{x+y-\alpha_d}{2}\right)\bigg\vert ,\\
\label{eq:Gamma_ecapture2}
\Gamma_{ue\to d\nu} (\mu_{\Delta}) 
&=& 
\frac{{G}^2T^8}{8\pi^5}\int_{-\infty}^\infty\!
dy\,\bar{y} g(\bar{y}) \! \int_0^{{\rm min}\{\alpha_d+y;\alpha_d +y +\Delta u\}}\!
dx \left[(\alpha_d +y +\Delta u)^2-x^2\right]\nonumber\\
&&\times \left[(\alpha_d+y)^2 -x^2\right]
\ln \bigg\vert 1+\exp\left(\frac{x+y-\alpha_d}{2}\right)\bigg\vert, 
\eea
which demonstrates that for $u_d=u_u$ the $d$-quark decay rate
vanishes in the ultrarelativistic limit according to
\eqref{eq:Gamma_ddecay2}.  For the $e$-capture rate, we find
\bea\label{eq:Gamma_ecapture3}
\Gamma_{ue\to d\nu} (\mu_{\Delta}) 
&=& 
\frac{{G}^2T^8}{8\pi^5}\int_{-\infty}^\infty\!
dy\,\bar{y} g(\bar{y}) \! \int_{-\alpha_d-y}^{0}\!
dx'\, \left[(\alpha_d+y +\Delta u)^2 -(x'+\alpha_d+y)^2\right] \nonumber\\
&&\times \left[-x'(2\alpha_d+2y+x')\right]
\ln \big\vert 1+\exp\left(x'/2+y\right)\big\vert,
\eea
where we made a variable change $x=x'+\alpha_d +y$ and made sure that
$\Delta u\neq 0$ in order to compute the derivative with respect to
$\mu_\Delta(\Delta u)$. If we now assume that $\beta$-equilibrium is
established with $\bar{y}=y$, and also $\Delta u =0$ we obtain from
Eq.~\eqref{eq:Gamma_ecapture3}
\bea\label{eq:Gamma_ecapture4}
\Gamma_{ue\to d\nu}  
&=& 
\frac{{G}^2T^8}{8\pi^5}\int_{-\infty}^\infty\!
dy\,{y} g({y}) \! \int_{-\alpha_d-y}^{0}\!
dx\, x^2 (2\alpha_d+2y+x)^2
\ln \big\vert 1+\exp\left(x/2+y\right)\big\vert
\nonumber\\
&\simeq & 
\frac{{G}^2T^8\alpha_d^2}{2\pi^5}\int_{-\infty}^\infty\!
dy\,{y} g({y}) \! \int_{-\infty}^{0}\!
dx\, x^2
\ln \left[ 1+\exp\left(x/2+y\right)\right],
\eea
where we dropped the smaller than $\alpha_d$ terms, and replaced the
lower boundary of integration by $-\infty$ (as the logarithm is
suppressed at small values of $x$).

To compute the bulk viscosity, consider next small departures from
$\beta$-equilibrium $\mu_{\Delta}\ll T$. Then, the net $u$-quark
production rate can be approximated as
$\Gamma_{d\to ue\bar\nu}-\Gamma_{ue\to d\nu}=\lambda_d\mu_{\Delta}$,
where $\lambda_d$ is the equilibration coefficient defined as
\bea \label{eq:lambda_def_d}
\lambda_d = 
-\frac{\partial\Gamma_{ue\to d\nu}}
{\partial\mu_{\Delta}}\bigg\vert_{\mu_{\Delta}=0}
&=& -\frac{{G}^2T^7 \alpha_d^2}{2\pi^5}
\bigg\{\int_{-\infty}^\infty\!
dy\, g(y)[1-y(1+g(y))] \! \int_{-\infty}^{0}\!
dx\, x^2 \ln \left[1+\exp\left(x/2+y\right)\right]\nonumber\\
&-&\int_{-\infty}^\infty\!
dy\, yg(y)\! \int_{-\infty}^{0}\!
dx\, x \ln \left[1+\exp\left(x/2+y\right)\right]
\bigg\}.
\eea
Note that when computing $\lambda_d$, we also take the derivative of
the terms containing $\Delta u$.

The integrals in Eqs.~\eqref{eq:Gamma_ecapture4}
and \eqref{eq:lambda_def_d} can be computed 
numerically, after which we obtain
\bea \label{eq:Gamma_lowT}
\Gamma_{ue\to d\nu}  
\simeq 0.38 {G}^2 p_{Fd}^2 T^6, \quad
\lambda_d 
\simeq 0.2{G}^2 p_{Fd}^2 T^5.
\eea
We observe that, as has been noted in other treatments of Urca
processes in quark matter \cite{Duncan1984ApJ,Burrows1980PhRvL}, the
$u\ e^- \to d$ rate contains an additional power of $T$ as compared to
the low-$T$ Urca reaction rates of nonrelativistic baryons. This is
because the $u$, $d$, and $e$ are all ultrarelativistic, so their
Fermi momenta are on the borderline between the Urca process being
allowed and being forbidden: the $u$ and $e^-$ momenta have to be
collinear in order to create a $d$. It is only the thermal blurring of
the Fermi surfaces that creates phase space for the process to occur.
Consequently, the product $(k \cdot p)$ in
Eq.~\eqref{eq:matrix_el_full} contributes an additional power of $T$
beyond the $T^5$ scaling that emerges when the phase space includes a
range of angles even in the low temperature limit.  It is worth noting
that the product $\left(k^{\prime} \cdot p^{\prime}\right)$ does not
introduce an additional power of $T$ because the angle between the
$d$-quark and neutrino momenta can be arbitrary, since the neutrino is
thermal.

Thus, the low-temperature direct Urca rates for light quarks (when the
isospin chemical shift is absent) differ qualitatively from those of
massive particles. For example, Urca reactions involving the $s$ quark
instead of the $d$ quark will have low-temperature rates similar to
the nucleonic Urca rates~\cite{Alford2023}
\bea\label{eq:Gamma_lowT_s}
\Gamma_{s\to ue\bar\nu}=\Gamma_{eu\to s\nu}= 0.0336
{G}^2 T^5 \mu_s^* \left[({p}_{Fu}+p_{Fe})^2-p_{Fs}^2\right]\theta(p_{Fe}+{p}_{Fu}-p_{Fs}).
\eea
The relevant coefficient $\lambda_s$ in the low-$T$ limit is given by
\bea\label{eq:lambda_lowT_s}
\lambda_s = \frac{17}{120\pi}{G}^2T^4 \mu_s^* 
\left[({p}_{Fu}+p_{Fe})^2-p_{Fs}^2\right].
\eea
In this limit we have also the relations $p_{Fe}=\mu_e$, $p_{Fu}\simeq
\mu_u^*=\mu_u-\omega_0$, $\mu_s^*=\mu_s-\phi_0=
\sqrt{p_{Fs}^2+m_s^2}$, therefore for the square brackets in
Eq.~\eqref{eq:lambda_lowT_s} we can write
$({p}_{Fu}+p_{Fe})^2-p_{Fs}^2\simeq
(\mu_u-\omega_0+\mu_e)^2-(\mu_s-\phi_0)^2+m_s^{*2}=(2\mu_s-\omega_0-\phi_0)(\phi_0-\omega_0)+m_s^{*2}$,
where we used the chemical equilibrium condition
$\mu_u+\mu_e=\mu_s$. Numerically, we find that $\omega_0, \phi_0\ll
m_s^*\simeq \mu_s^*$, therefore, the square brackets in
Eq.~\eqref{eq:lambda_lowT_s} can be approximated as $m_s^{*2}$, which
leads to the simple expressions
\bea\label{eq:Gamma_lowT_s1}
\Gamma_{s\to ue\bar\nu}=\Gamma_{eu\to s\nu}= 0.0336
{G}^2 \mu_s^* m_s^{*2} T^5,\quad 
\lambda_s 
\simeq 0.03 {G}_F^2 \sin^2\theta_c\mu_s^* m_s^{*2}T^4.
\eea
To estimate the rate of non-leptonic processes~\eqref{eq:non-leptonic}
we will use below the low-temperature expression~\cite{Madsen1993}
\bea\label{eq:lambda_non_lep}
\lambda_{\rm non-lep} =\frac{64}{5\pi^3} {G}_F^2 \sin^2\theta_c\cos^2\theta_c\mu_d^{*5} T^2.
\eea

\section{Bulk viscosity of $udse$ matter}
\label{sec:bulk_visc}

In this section, we will derive the bulk viscosity that arises from
processes \eqref{eq:d_decay}-\eqref{eq:e_capture_s} in $udse$ matter,
i.e. matter consisting of $u,d,s$ quarks and electrons,
with paired red-green light quarks; the unpaired excitations are blue
light quarks, strange quarks of all colors, and leptons (electrons,
and muons if energetically favored). Consider small-amplitude density oscillations with a
frequency $\omega$. Separating the oscillating parts from the static
equilibrium values of particle densities, we can write
$n_j(t)= n_{j0}+\delta n_j(t)$, where
$\delta n_j(t)\sim e^{i\omega t}$, and $j=\{d,u,e,s\}$ labels the
particles.

Oscillations drive the system out of chemical equilibrium, leading to
nonzero chemical imbalance
$\mu_{\Delta}= \delta\mu_d-\delta\mu_u-\delta\mu_e$ in the case of
$ude$ matter.  To include strange quarks, note that the non-leptonic
reaction $d+u\leftrightarrow s+u$ proceeds much faster than the Urca
processes; therefore, the relation $\mu_d=\mu_s$ always holds, and the
shift in chemical potentials is given by
$\mu_{\Delta}=\delta\mu_d-\delta\mu_u-\delta\mu_e=\delta\mu_s-\delta\mu_u-\delta\mu_e$,
which can be written as
\bea\label{eq:delta_mu_1} 
\mu_{\Delta} = A_d \delta n_d +A_s\delta n_s  -A_u \delta n_u -A_e
\delta n_e,
\eea 
where the susceptibilities $A_j$ and are given as 
\bea\label{eq:A_d_fast} 
A_d = A_{dd}-A_{ud},\quad
A_u = A_{uu}-A_{du},\quad 
A_s= A_{ds}-A_{us},\
A_e = A_{ee}, 
\eea
where $A_{ij} = {\partial \mu_i}/{\partial n_j},$ and the derivatives
are computed in the beta-equilibrium state. Note that off-diagonal
elements $i\neq j$ do not vanish because of strong interactions
between quarks.

If the weak processes were switched off, then the number of all
particle species would be conserved separately, which implies
\bea\label{eq:cont_j}
\frac{\partial}{\partial t} \delta {n}^0_j(t)+ \theta n_{j0} =0
, \quad \delta {n}^0_j(t) = -\frac{\theta}{i\omega}\, n_{j0},
\eea 
where $\theta=\partial_i v^i$ is the fluid velocity divergence. Once
the weak reactions are switched on, there is a net production of
particles that should be the neutrino production rates by quarks,
given by
\bea\label{eq:lambda_def_d1}
\Gamma_{d\to ue\bar{\nu}}-\Gamma_{ue\to d\nu} &=& \lambda_d\mu_\Delta,\\
\Gamma_{s\to ue\bar{\nu}}-\Gamma_{ue\to s\nu} &=& \lambda_s\mu_\Delta,
\eea
which define the equilibration coefficients $\lambda_{d}$ and $\lambda_{s}$.
Therefore, the rate equations for each fermion species can be written in this case as 
\bea\label{eq:cont_d_fast}
\frac{\partial}{\partial t}\delta n_d &=& 
-\theta  n_{d0} -\lambda_d \mu_\Delta -I_{ud\rightarrow us},\\
\label{eq:cont_s_fast}
\frac{\partial}{\partial t}\delta n_s &=& 
-\theta  n_{s0} -\lambda_s \mu_\Delta +I_{ud\rightarrow us},\\
\label{eq:cont_u_fast}
\frac{\partial}{\partial t}\delta n_u &=& 
-\theta n_{u 0} +(\lambda_d+\lambda_s)\mu_\Delta,\\
\label{eq:cont_e_fast}
\frac{\partial}{\partial t}\delta n_e &=& 
-\theta n_{e 0} +(\lambda_d+\lambda_s)\mu_\Delta.
\eea 
where $I_{ud\rightarrow us}$ denotes the rate of the non-leptonic
reaction $d + u \rightarrow s + u$, which is driven by a nearly
vanishing chemical potential difference,
$\delta\mu_{d}-\delta\mu_{s}\ll \mu_\Delta$. Despite its small
magnitude, this shift cannot be neglected because the corresponding
$\lambda$-coefficient may be very large; see
Ref.~\cite{Jones2001PhRvD} for a discussion of this point.

Among the resulting balance equations, only one is independent due to
the presence of three constraints: charge neutrality (both color and
electric) and baryon number conservation. These constraints can be
expressed in the form:
\bea\label{eq:baryon_cons}
\tilde{n}_u+\tilde{n}_d+\tilde{n}_s = 2(n_u+ n_d+n_s)&=& 2n_b,\\
\label{eq:charge_neutrality}
\frac{2}{3}(n_u+\tilde{n}_u) -\frac{1}{3}(n_d+n_s+\tilde{n}_d+\tilde{n}_s) &=& 
n_e +n_\mu = n_u+\tilde{n}_u -n_b,
\eea 
where we denote with $n_i$ the densities of only blue quarks, and with
$\tilde{n}_i$ -- the summed densities of red and green quarks, and
$n_b$ is the baryon density. Then we find
\bea\label{eq:n_relations_n1}
\delta n_d +\delta n_s &=& \delta n_b-\delta n_u,\qquad
\delta \tilde{n}_d +\delta \tilde{n}_s = 2\delta n_b-\delta \tilde{n}_u,\\
\label{eq:n_relations_charge1}
\qquad \delta n_e + \delta n_\mu &=&
\delta n_u+ \delta \tilde{n}_u-\delta n_b =\delta \tilde{n}_u -\delta n_d-\delta n_s,\\
\label{eq:n_relations_mu1}
\mu_\Delta
 &=&
 \left(A_d+A_e\right)\delta n_b -A_1\delta n_u
 -A_e(\delta \tilde{n}_u-\delta n_\mu)
 +(A_s-A_d) \delta n_s,
\eea 
where $A_1=A_u+A_d+A_e$.

Substituting Eq.~\eqref{eq:n_relations_mu1} into
Eqs.~\eqref{eq:cont_u_fast}, assuming $\delta n_j\sim e^{i\omega t}$
and using Eq.~\eqref{eq:cont_j} for $n_b$, $\tilde{n}_u$, and $n_\mu$
(the paired quarks and muons do not participate in reactions), we
obtain
\bea\label{eq:cont_u1}
i\omega\delta n_u
 &=& -\theta n_{u 0} +\lambda(A_d+A_e)
 \delta n_b -\lambda A_1\delta n_u -\lambda A_e(\delta \tilde{n}_u  -\delta n_\mu)
 +\lambda(A_s-A_d) \delta n_s,
\eea 
where $\lambda=\lambda_d+\lambda_s$ is the summed rate of the
$u$-quark production by Eqs.~\eqref{eq:d_decay}--\eqref{eq:e_capture_s}.
To eliminate $\delta n_s$ from this equation, we use
 the condition of chemical equilibrium with
respect to non-leptonic reaction 
\bea\label{eq:delta_mu_ds}
0 &=& \delta\mu_d -\delta\mu_s
=(A_{dd}-A_{sd})\delta n_d +(A_{du}-A_{su})\delta n_{u}
+(A_{ds}-A_{ss}) \delta n_s \nonumber\\
&=&A_B \delta n_b + A_U\delta n_u +A_S\delta n_s,
\eea
which can be solved for $\delta n_s$, and 
where we introduced shorthand notations
\bea
A_B=  A_{dd}-A_{sd}, \quad \quad
A_U =  A_{du}-A_{su}-A_{dd}+A_{sd},\quad
A_S = A_{ds}-A_{ss}-A_{dd}+A_{sd}.
\eea
Substituting $\delta n_s$ from Eq.~\eqref{eq:delta_mu_ds} back into
Eq.~\eqref{eq:cont_u1} and using Eq.~\eqref{eq:cont_j} gives
\bea\label{eq:delta_u_fast}
\delta n_u
&=& -\frac{i\omega n_{u0} +\lambda (B+A_e) n_{B0}
  -\lambda A_e (\tilde{n}_{u0}-n_{\mu0})}
{i\omega+\lambda A}\frac{\theta}{i\omega}.
\eea 
Subtracting from this expression
$\delta n^{0}_u = -{\theta n_{u0}}/{i\omega}$ we obtain the nonequilibrium shift
\bea\label{eq:delta_u_prime_fast1}
\delta n'_u &=&\delta n_u-\delta n^{0}_u
= -\frac{C}
{i\omega+\lambda A}\frac{\theta\lambda}{i\omega}, \\
C &\equiv& B (n_{d0}+n_{s0})-(A-B-A_e) n_{u0} -A_e n_{e0},\\
A &\equiv& A_1 +\frac{(A_s-A_d)A_U}{A_S}, \quad
B \equiv A_d -\frac{A_U A_B}{A_S}.
\eea 
Solving Eq.~\eqref{eq:delta_mu_ds} for $\delta n'_s$ (recall that the
non-equilibrium shifts of $n_b$, $n_\mu$ and the paired quarks are
zero) we find 
\bea\label{eq:delta_s_prime_fast}
\delta n'_s = -\frac{A_U}{A_S}\delta n'_u.
\eea 
Then the bulk viscous pressure will be given by (using short-hand
notation $c_j={\partial p}/{\partial n_j}$)
\bea\label{eq:Pi_fast}
\Pi &=& \sum_j
\frac{\partial p}{\partial n_j}\delta n'_j =
\left[(c_u-c_d +c_e)+(c_d-c_s) \frac{A_U}{A_S}\right]\delta n'_u,
\eea 
where we used Eqs.~\eqref{eq:n_relations_n1} and
\eqref{eq:n_relations_charge1}.  Assuming isothermal perturbations to
compute the pressure derivatives, and further using
Eq.~\eqref{eq:delta_u_prime_fast1} and the symmetry relation
$A_{ij}=A_{ji}$ we find
\bea\label{eq:Pi_fast1}
\Pi = -C\delta n'_u = \frac{C^2\lambda}
{i\omega+\lambda A}\frac{\theta}{i\omega}.
\eea 
The bulk viscosity is the real part of $-\Pi/\theta$ and is thus defined as 
\bea\label{eq:zeta}
\zeta = \frac{C^2}{A}\frac{\gamma}{\omega^2+\gamma^2},
\eea
which has the classic resonant form depending on two quantities: the
susceptibility prefactor $C^2/A$, which depends only on the EoS, and
the relaxation rate $\gamma=\lambda A$, which depends on the EoS and
the microscopic interaction rates.  In the case where the non-diagonal
susceptibilities can be neglected, the quantities $A$ and $C$ are
given by
\bea\label{eq:A_non_int}
A &=&  \frac{A_dA_s}{A_d+A_s} +A_u+A_e 
= -\dfrac{1}{n_b}
 \dfrac{\partial \mu_\Delta}{\partial x_u} \bigg\vert_{n_b} ,\\
\label{eq:C_non_int}
C &=& \frac{A_dA_s}{A_s+A_d}(n_{d0}+n_{s0}) 
-A_u n_{u0}-A_e n_{e0}
= n_b\dfrac{\partial \mu_\Delta}{\partial n_b} \bigg\vert_{\!x_u},
\eea 
where we redefined $A_i={\partial \mu_i}/{\partial n_i}$ which are computed in chemical
equilibrium.  If we neglect the contribution from $s$-quarks, then
$n_{s0}\to 0$, $A_s\to \infty$, and we find the appropriate quantities
for the $ude$ quark matter \bea A =A_u+A_d+A_e, \quad C = n_{d0} A_d-
n_{u0}A_u - n_{e0}A_e.
\eea
We note that $A$ can be interpreted as the
beta-disequilibrium–$u$-quark-fraction susceptibility: it quantifies
how the out-of-beta-equilibrium chemical potential responds to a
change in the $u$-quark fraction. Similarly, $C$ is the
beta-disequilibrium–baryon-density susceptibility: it characterizes
the response of the out-of-beta-equilibrium chemical potential to a
change in the baryon density, while keeping the particle fractions
fixed.

\section{Numerical results}
\label{sec:num_results}

\subsection{Equilibration coefficients and Urca rates}

\begin{figure}[bt]
\begin{center}
\includegraphics[width=17cm]{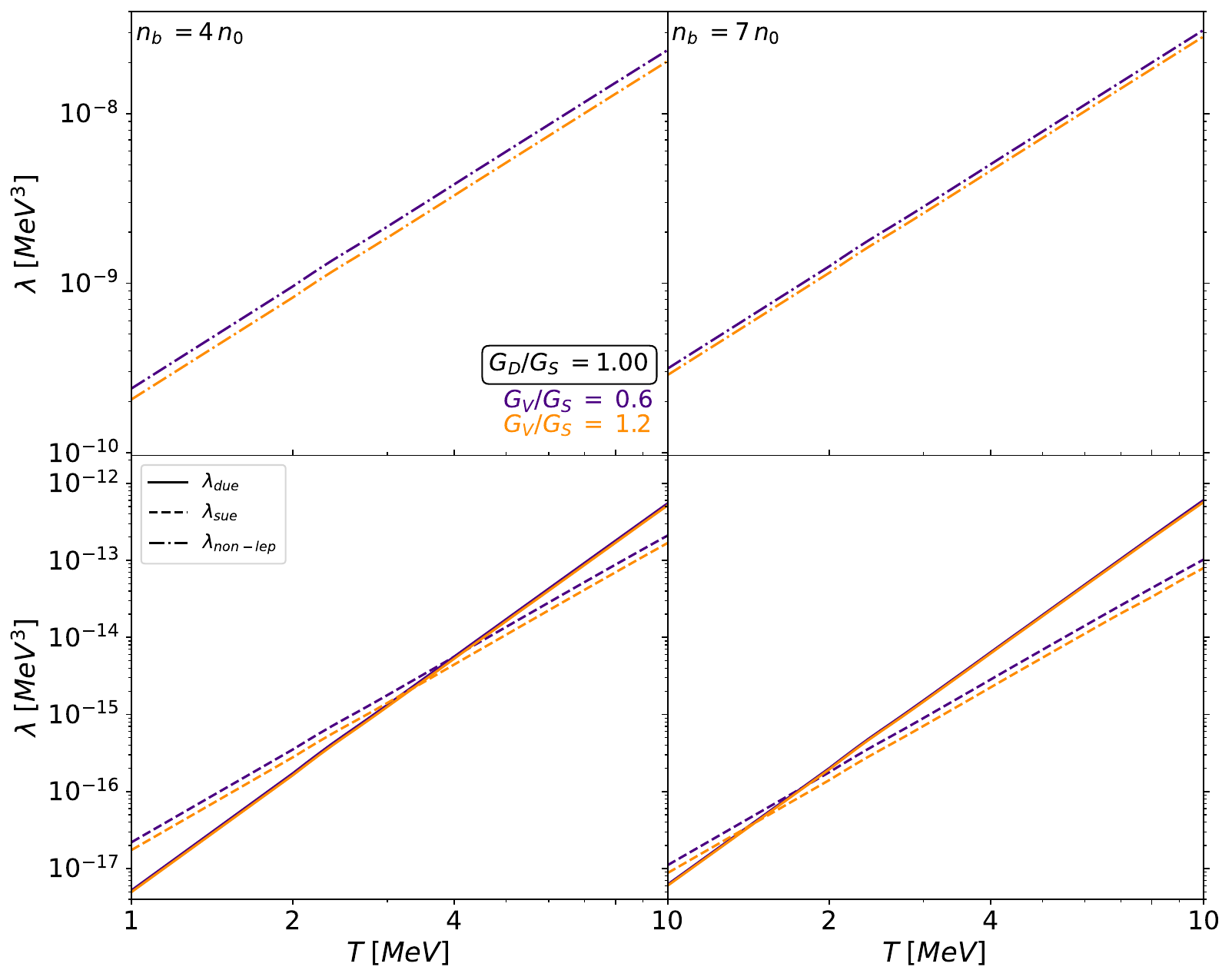}
\end{center}
\caption{The equilibration coefficients $\lambda$ of Urca and
non-leptonic processes as functions of temperature for fixed
values of number density and various fixed values of vector and
 diquark couplings.}
\label{fig:lambda}
\end{figure}
As expected, the non-leptonic processes described by
Eq.~\eqref{eq:non-leptonic} occur at significantly higher rates than
the Urca processes. This is evident in Fig.~\ref{fig:lambda}, which
shows the equilibration coefficients for both the Urca processes
[Eqs.~\eqref{eq:d_decay}--\eqref{eq:e_capture_s}] and the non-leptonic
processes [Eq.~\eqref{eq:non-leptonic}]. Notably, in the temperature
range $1 \leq T \leq 10$~MeV the rates of the $d$-Urca and $s$-Urca
processes are comparable for the matter composition predicted by the
vector-enhanced NJL model. As a result, the bulk viscosity associated
with the non-leptonic channels is expected to peak at much lower
temperatures, well below the relevant range for BNS mergers, whose
oscillation frequencies lie in the kilohertz regime. This observation
supports our assumption that, in the temperature range
$1 \leq T \leq 10$~MeV, the bulk viscosity can be reliably calculated
from the Urca processes alone, under the additional condition
$\mu_d=\mu_s$. The effects of diquark and vector couplings enter the
equilibration coefficients through the composition (chemical
potentials, pairing gap, etc.) of the participating particles. It is
seen that the combined effect on the equilibration coefficients is to
decrease both non-leptonic and Urca processes with increasing vector
coupling. The changes associated with variations of the diquark
coupling are insignificant.

Analytical expressions for $\lambda_d$ and $\lambda_s$ coefficients
were derived in the low-temperature limit, see
Eqs.~\eqref{eq:Gamma_lowT} and \eqref{eq:Gamma_lowT_s1} under the
assumptions that $T\ll \mu_i$, $m_d, m_s\to 0$ and
$\omega_0-\phi_0\ll m_s$. They are accurate to within a few percent
when compared to numerical results that do not rely on these
approximations and provide insight into the scaling of these
coefficients with various parameters. In particular, it is seen that
$\lambda_s(T)$ is suppressed by a factor of $\sin^2\theta_c=0.05$
compared to $\lambda_d(T)$ which has $\cos^2\theta_c=0.95$; it also
has a smaller numerical prefactor from the phase space
integration. Due to the influence of interactions on the chemical
potentials of light quarks mediated by the isoscalar $\omega$-field,
the $\beta$-equilibrium for massless particles requires that
$p_{F d}=p_{F u}+p_{F e}$. This condition implies that the direct Urca
process is only thermally allowed. In contrast, the $s$-Urca channels
are kinematically open, with a substantial available energy range
given by $p_{F u}+p_{F e}-p_{F s} \geq 70~\mathrm{MeV}$, primarily due
to the large mass of the strange
quark~\cite{Duncan1983ApJ,Duncan1984ApJ}.  We previously discussed the
additional power of $T$ arising from the matrix element involving the
four-product of the $u$-quark and electron momenta, which are
massless. In contrast, the phase-space contributes the standard $T^5$
scaling, which would be the only temperature-dependent component if
all particles were massive (implicit weak dependence of masses and
other thermodynamic parameters is understood).  As a consequence, the
difference in temperature scaling leads to the intersection of the
$\lambda_d(T)$ and $\lambda_s(T)$ functions at a temperature in the
MeV range, as seen in Fig.~\ref{fig:lambda}.
\begin{figure}[bt]
\begin{center}
\includegraphics[width=12cm]{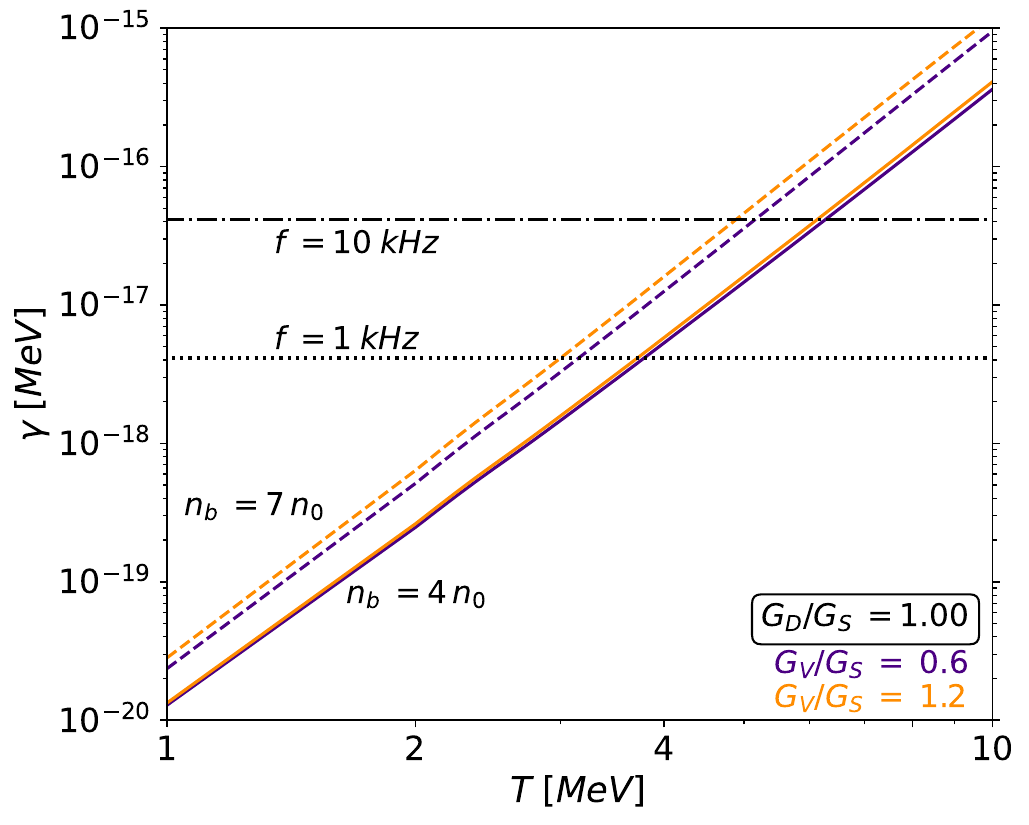}
\end{center}
\caption{The $\gamma$ parameter  as a function of temperature for two
  values of number density and various fixed values of vector
  and diquark couplings. }\label{fig:gamma}
\end{figure}

We show in Fig.~\ref{fig:gamma} the temperature dependence of the
total $\beta$-relaxation rate $\gamma$ due to the Urca
processes~\eqref{eq:d_decay}--\eqref{eq:e_capture_s} as a function of
temperature for different fixed values of density, diquark, and vector
couplings. It follows a power law scaling $\gamma\propto T^{4.5}$ in
the temperature range $1\leq T\leq 10$~MeV. A larger vector coupling
enhances the rate $\gamma$ slightly, this enhancement being more
pronounced at higher density. There are no significant changes with
the diquark coupling in the limit $T\ll \Delta$. Due to the Lorentzian
structure of bulk viscosity in the frequency domain, see
Eq.~\eqref{eq:zeta}, its maximum is located at the ``resonance''
temperature determined by equating $\gamma$ to the characteristic
angular frequency $\omega=2\pi f$.  For illustration, we consider two
representative frequencies. At $f=1$~kHz, the intersection occurs near
$T \simeq 3-4$ MeV, while for $f=10$ kHz, it shifts to approximately
$T \simeq 5-6$ MeV.

\subsection{Bulk viscosity and damping timescales}

\begin{figure}[h!]
\begin{center}
\includegraphics[width=17cm]{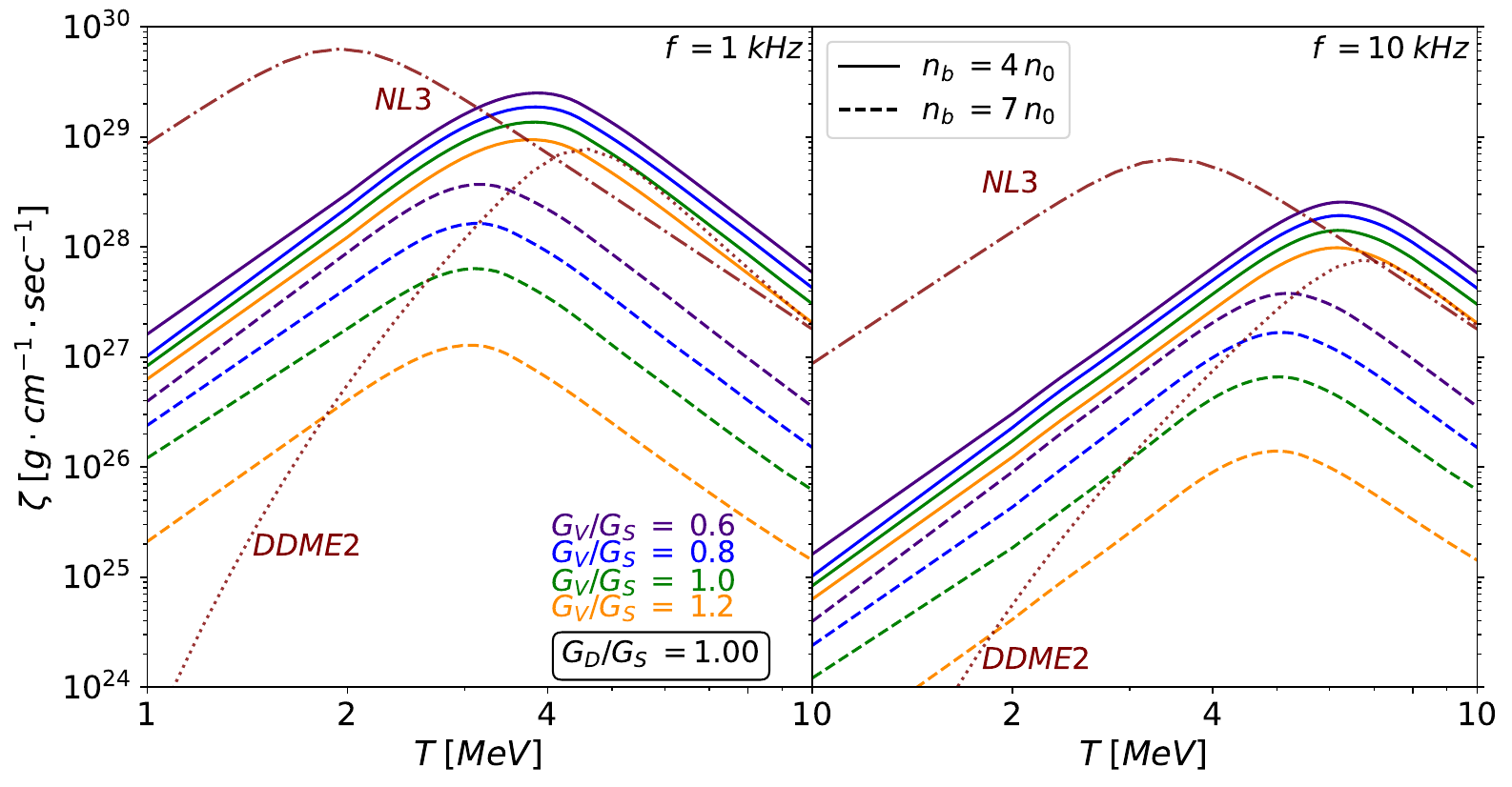}
\end{center}
\caption{Bulk viscosity as a function of temperature for $n_b= 4n_0$
  (solid lines) and $n_b=7 n_0$ (dashed lines) and various fixed
  values of vector and diquark
  couplings; here $n_0$ is the nuclear saturation density. For
  comparison, we also show the bulk viscosity of
  neutrino-transparent nucleonic matter, as was computed in
  Ref.~\cite{Alford2023} for models DDME2 (dotted line) and NL3
  (dash-dotted line) for $n_b/n_0=4$. 
\label{fig:zeta}
}
\end{figure}

Next, we turn to the main quantity of interest of this study - the
bulk viscosity of neutrino-transparent quark matter in the 2SC phase.
Figure~\ref{fig:zeta} shows the temperature dependence of bulk
viscosity for density oscillations at two frequencies $f=1$~kHz and
$f=10$~kHz and at two baryon densities $n_b= 4n_0$ and $n_b=7 n_0$.
As a key feature, the figure also shows the influence of the variation
of the vector coupling $G_V/G_S$.  The diquark coupling is set to
$G_D/G_S=1$; results for $G_D/G_S=1.25$ were also calculated but are
not shown because they are almost identical to those for
$G_D/G_S=1$. This is understandable, since for either of these values
of $G_D$, the gap in the spectrum of the paired quarks is much greater
than the temperature, so they are frozen out.

The curves of bulk viscosity as a function of temperature have the
Lorentzian shape expected from \eqref{eq:zeta}. For density
oscillations of angular frequency $\omega$, the maximum is reached
when $\gamma(T)=\omega$, so its location along the $T$ axis is
determined by the temperature dependence of the relaxation rate
$\gamma$.

As density increases, the maximum slowly moves to lower temperatures
because the relaxation rate $\gamma$ rises with density (as the
increase in phase space near the Fermi surfaces leads to faster Urca
rates), so $\gamma(T)=\omega$ is achieved at lower temperatures. The
leftward shift in the curve is small because $\gamma$ rises quickly,
roughly as $T^{4.5}$,so it only takes a small reduction in $T$ to
compensate for the effect of rising density. This temperature
dependence differs from the $T^4$ scaling seen in
Ref.~\cite{Hernandez2024} because it is a combination of
$\lambda_s\sim T^4$ scaling for the $u\to s\ e^-\ \bar\nu_e$ channel
(Eq.~\eqref{eq:Gamma_lowT_s}) and $\lambda_d\sim T^5$ for the
$u\to d\ e^-\ \bar\nu_e$ Urca channel (Eq.~\eqref{eq:Gamma_lowT}).
The position of the maximum is even less affected by changes in the
vector coupling $G_V$, since it has only a small effect on the Urca
rates, see Fig.~\ref{fig:lambda}.  We also find that the effect of
changing $G_D$ is quite small, therefore, we do not show this case
explicitly.

The overall scale of the bulk viscosity curve (e.g., the value
attained at the maximum) is controlled by the prefactor $C^2/A$ in
Eq.~\eqref{eq:zeta}. The susceptibilities $C$ and $A$ are affected by
the density and the couplings.  As density increases, the overall
scale decreases considerably, mainly due to a decrease in the $C$
susceptibility of 2SC matter. We also see a noticeable effect of the
couplings.  At lower density $n_b/n_0 =4$, doubling the vector
coupling reduces the bulk viscosity at all temperatures by a factor of
3. At higher density $n_b/n_0 =7$, doubling the vector coupling
reduces the bulk viscosity by more than an order of magnitude. Without
explicitly showing in the figure, let us point out that increasing the
magnitude of the diquark coupling $G_D/G_S$ leads to a further
decrease in the bulk viscosity, which is more pronounced for a larger
density $n_b/n_0 =7$. The magnitude of the decrease is about a factor
of two.  The temperature dependence of the bulk viscosity $\zeta(T)$
is seen to be self-similar for the curves shown. From
Eq.~\eqref{eq:zeta} the rate goes as $\gamma$ well below the maximum
and as $1/\gamma$ well above the maximum, so from the scaling given in
the previous paragraph one would expect $\zeta \propto T^{4.5}$ and
$\zeta \propto T^{-4.5}$ respectively. However, there is also a small
temperature dependence from the susceptibilities, so the scaling is
closer to $\zeta(T) \propto T^{-4.2}$ on the descending side.

We also reiterate that, as noticed in Ref.~\cite{Alford2024PhRvDLett},
including/excluding the strange quarks increases/decreases the total
relaxation rate by a factor of 1 to 2, resulting in only a slight
shift of the viscosity peak toward lower/higher temperatures.

For comparison, we also show the bulk viscosity of
neutrino-transparent nucleonic matter, as calculated in
Ref.~\cite{Alford2023}, using the DDME2 (dotted line) and NL3
(dash-dotted line) models at a baryon density of $n_b/n_0=4$. The key
difference between these two models lies in the behavior of the direct
Urca process at low temperatures: it is allowed for the NL3 model but
remains blocked for DDME2 at this density. Consequently, the bulk
viscosities predicted by the two models differ markedly at
temperatures $T\leq 4$~MeV, but are rather similar at $T\geq
4$~MeV. In this temperature range, the bulk viscosity of 2SC quark
matter is similar to that of nucleonic matter, and variations in the
vector or diquark couplings do not significantly affect this outcome.

At lower temperatures, the bulk viscosity of 2SC quark matter falls
between the values predicted by the two nucleonic models, regardless
of the specific values of the vector and diquark couplings. This
behavior can be attributed to the $u$-quark fraction being slightly
below, but very close to the threshold for the direct Urca processes
described in Eqs.~\eqref{eq:d_decay} and
\eqref{eq:e_capture_d}. Specifically, the difference in Fermi momenta
between initial-state and final-state particles,
$p_{Fd} - p_{Fu} - p_{Fe}$, is less than 1\,MeV for densities in the
range $4 \leq n_b/n_0 \leq 7$. As a result, thermal smearing at
temperatures as low as $T \geq 1$\,MeV is sufficient to provide the
phase space required for light-quark Urca processes to occur.

Since the bulk viscosity arises from semi-leptonic weak interactions
involving the unpaired blue quarks, one would expect similar behavior
in unpaired quark matter, specifically a peak in the bulk viscosity at
MeV-scale temperatures. This feature is indeed observed in
Ref.~\cite{Alford:2006gy} (Fig.~8), Ref.~\cite{Sad2007b} (Fig.~3) and
in Ref.~\cite{Hernandez2024} (Fig.~4), although the maximum viscosity
reported in those studies is smaller due to their assumption of a
lighter strange quark.

\begin{figure}[bt]
\begin{center}
\includegraphics[width=16cm]{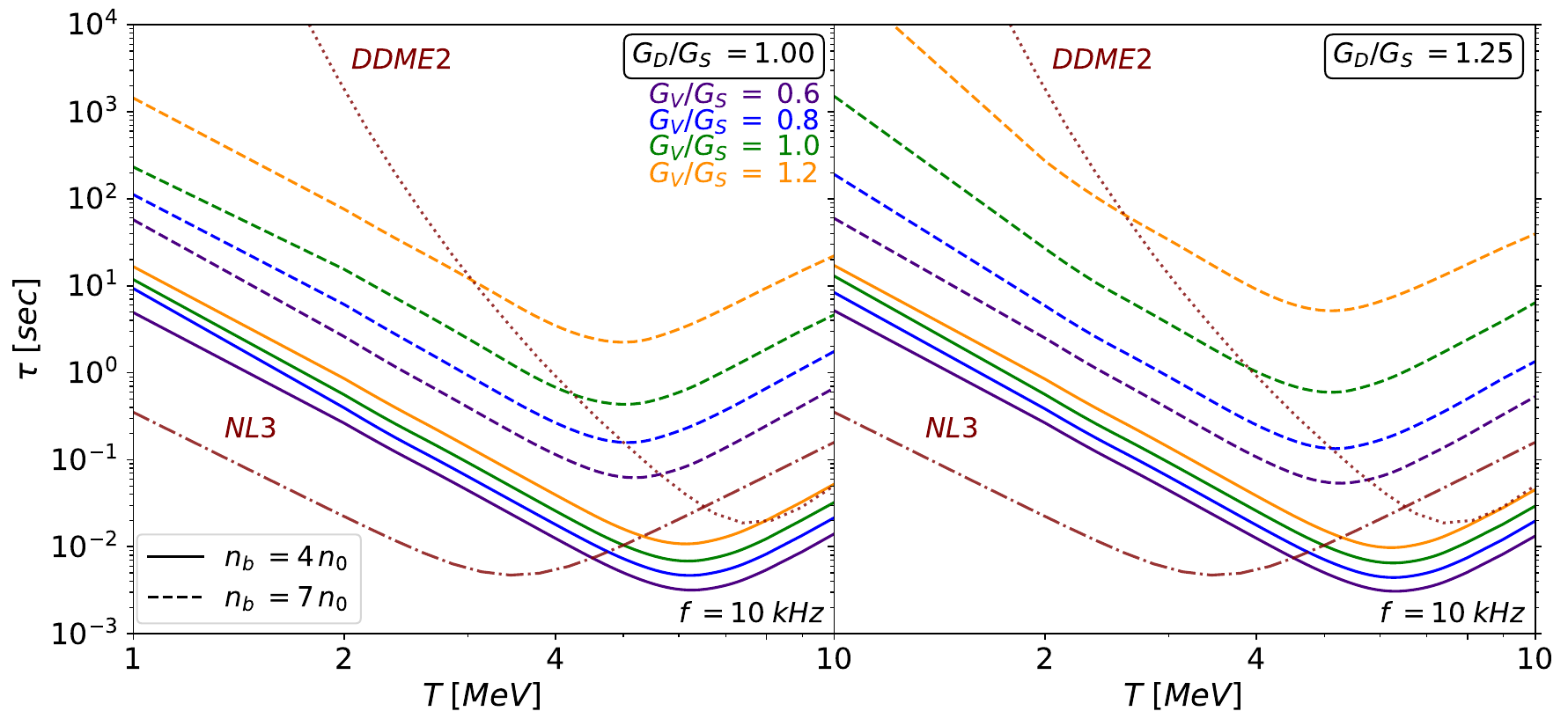}\\%
\includegraphics[width=16cm]{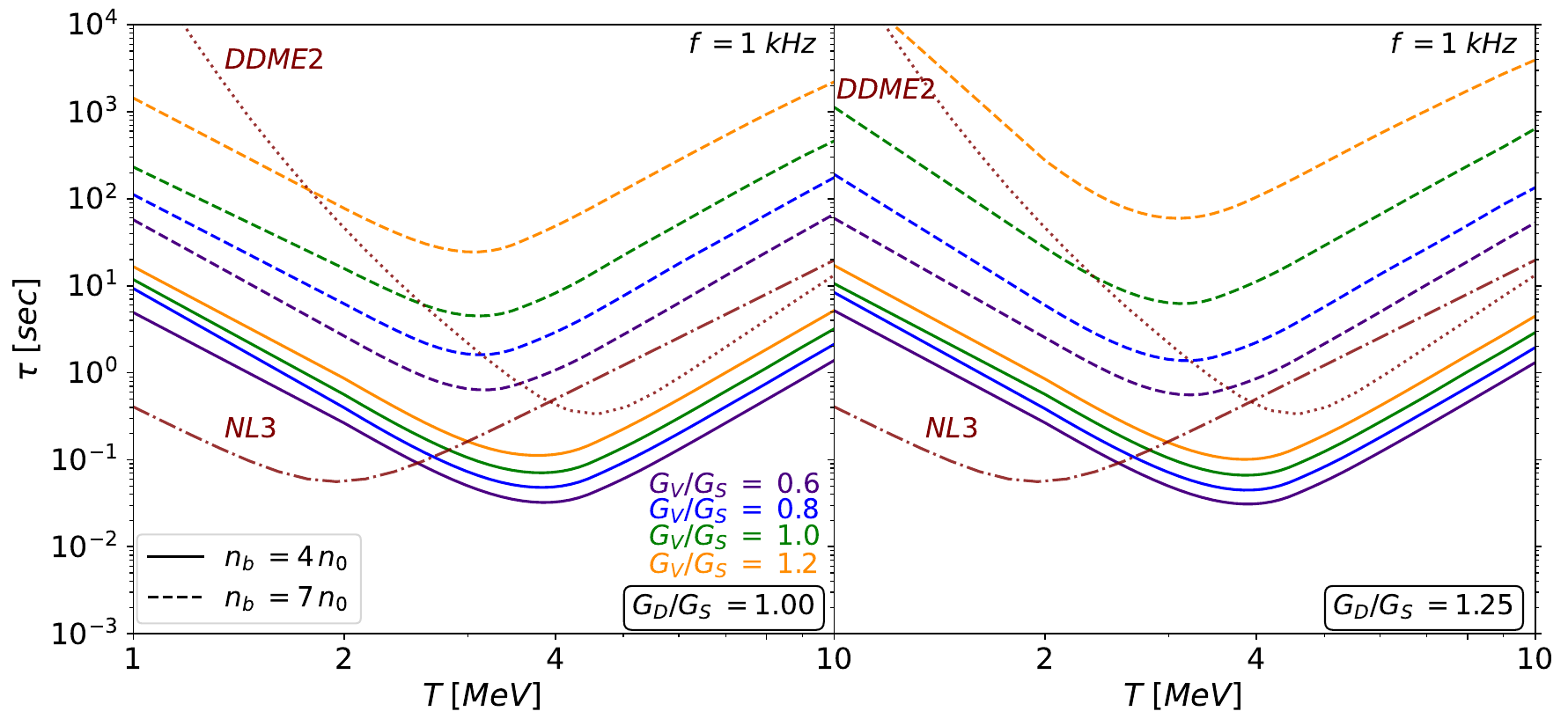}\\%
\end{center}
\caption{Dependence of bulk viscous damping timescales on temperature
  for two values of number density and various fixed values of vector
  and diquark couplings. For comparison, we also show the damping
  timescales of neutrino-transparent nucleonic matter as was computed
  in Ref.~\cite{Alford2023} for models DDME2 (dotted line) and NL3
  (dashed line) for $n_b/n_0=4$. }
\label{fig:tau}
\end{figure}

In the aftermath of a BNS merger, the resulting remnant undergoes
rapid and large-amplitude density oscillations, driven by strong
differential rotation, internal thermal gradients, and potentially
turbulent motion.  The remnant's temperature is highly variable over
its density range, and some regions may pass through or remain within
the temperature window where the bulk viscous damping is especially
efficient. If so, the bulk viscosity can significantly damp
oscillation modes, converting mechanical energy into heat (or radiated
neutrinos) and altering the thermal evolution of the remnant.  The
characteristic time scale of damping of local oscillations (on the
hydrodynamic scales characterized by fixed density, temperature,
entropy, etc.) is given by~\cite{Alford2018a,Alford2019a,Alford2020}
\bea\label{eq:damping_time}
\tau =\frac{1}{9}\frac{Kn_b}{\omega^2\zeta}, \quad K =9 \frac{\partial P}{\partial n_b},
\eea 
where $K$ is the incompressibility and $P$ is the pressure.  Using our
results for the bulk viscosity, we have estimated the damping
timescale from Eq.~\eqref{eq:damping_time}. The results are 
shown in Fig.~\ref{fig:tau}.  As
$\tau \propto \zeta^{-1}$, we see an inversion in the temperature
dependence in the damping time scale, which implies that it is
shortest at the resonant maximum of the bulk viscosity.  We have
verified that the compressibility $K$ in \eqref{eq:damping_time} is
insensitive to the temperature within the relevant range
$1\le T\le 10$~MeV.  Turning to the dependence of the damping
timescale $\tau$ on the diquark and vector couplings, we find that its
behavior is determined by that of the bulk viscosity $\zeta$. At lower
density $(n_b/n_0=4)$, 
varying the vector coupling by a factor of 2
changes the maximum value of $\tau$ by roughly a factor of 3, with the
shortest damping times corresponding to the smallest values of the
vector and diquark couplings. At higher density ($n_b/n_0=7$), the
damping timescale becomes even more sensitive to the vector coupling:
varying the coupling by a factor of 2 changes $\tau$ by more than an
order of magnitude. While a smaller diquark coupling $G_D / G_S$ also
leads to shorter damping times for a fixed vector coupling, the
variation it introduces is, in comparison, not large.

Overall, we find that the damping timescales range from a few
milliseconds up to several hundred milliseconds, which are comparable
to the characteristic timescales of the short-term evolution of the
post-merger remnant. For longer-lived remnants, which can survive for
several seconds or more, bulk viscous dissipation may play an even
more significant role in their dynamical evolution.

Furthermore, a comparison with the results obtained for nucleonic
matter reveals that the damping timescales in 2SC quark matter are
remarkably similar. This suggests that, within the framework of our
NJL model, it would be challenging to distinguish 2SC quark matter
from nuclear matter based solely on their bulk viscous dissipation
properties in the temperature range of 1-10 MeV.

\section{Conclusions}
\label{sec:conclusions}

Expanding on our previous work~\cite{Alford2024PhRvDLett}, we have
studied the effects of variations in diquark and vector couplings --
specifically $G_D/G_S=1,\, 1.25$ and $G_V/G_S=0.6,\,0.8,\,1.0,\,1.2$
-- on the input parameters used to compute the weak interaction rates
and the bulk viscosity of the 2SC phase of finite-temperature quark
matter within the vector-enhanced NJL model. We focus on the low
temperature range $1 \leq T \leq 10 ~\mathrm{MeV}$ and the baryon
density range $4 n_0 \leq n_b\leq 7 n_0$, within which neutrinos are
expected to be free-streaming.

The primary effect of varying the vector coupling is a shift in the
quark chemical potentials, which affects material properties such as
bulk viscosity in two ways. Firstly, it changes the relevant static
susceptibilities in the EoS; secondly, it changes
flavor-changing Urca rates by altering the available phase space. In
contrast, changing the diquark coupling mainly affects the degree of
suppression of red-green quark contributions by changing the pairing
gap. However, because red-green quarks are indirectly coupled to the
unpaired blue-quark sector through $\beta$-equilibrium and charge
neutrality conditions, their influence on the overall thermodynamics
is nontrivial and cannot simply be neglected.

Quantitatively, we find that varying the vector coupling by a factor
of 2 changes the bulk viscosity and corresponding damping timescale by
a factor of 3--20 at densities from $4n_0$ to $7n_0$ This sensitivity
primarily arises from the $C$ susceptibility of 2SC matter, see
Eq.~\eqref{eq:C_non_int}, with a smaller contribution from
modifications to the weak interaction rates. In comparison, changes in
the diquark coupling have a more limited impact.  The bulk viscosity
and damping time for 2SC quark matter closely resemble those of
nucleonic matter \cite{Alford2023}, making it difficult to distinguish
these states via their bulk viscous behavior.

It is worth noting that at temperatures below the range studied here,
$T\lesssim 0.1\,\MeV$, one can no longer rely on our assumption that
the non-leptonic flavor equilibration process can be treated as
instantaneous. When its finite rate is taken into account, the bulk
viscosity shows a second peak at $T\sim 0.05$\,MeV
\cite{Alford:2006gy,Rojas2024,Hernandez2024}. This peak is absent in
nuclear matter. The peak we observe at temperatures in the MeV range,
arising from semi-leptonic processes, is also seen in
Ref.~\cite{Hernandez2024} (Fig.~4) for their heaviest strange quark,
$m_s\geq 200$~MeV. In our model the strange quark is even heavier,
$300\leq m_s\leq 400$\,MeV so the semileptonic peak can be treated as
a separate feature. Note that the picture may be quantitatively
different in other partly-paired phases. For example, Urca rates in
the gapless 2SC phase, as computed in Ref.~\cite{Jaikumar2006}, may
allow the red-green quarks to contribute to the Urca rates at levels
comparable to blue quarks. Furthermore, as in our setup, blue quarks
are unpaired, one could anticipate that semi-leptonic rates for
unpaired quark matter may be similar to those for 2SC matter studied
here. By extrapolation, this suggests that the unpaired quark matter
at $T\approx 1$ to $10$~MeV may also be difficult to distinguish from
nuclear matter, as was found for the 2SC phase.

\section*{Acknowledgments}

This work was partly supported by the U.S. Department of Energy,
Office of Science, Office of Nuclear Physics under Award
No. DE-FG02-05ER41375 (M.~A.), the Higher Education and Science
Committee (HESC) of the Republic of Armenia through “Remote
Laboratory” program, Grant No. 24RL-1C010 (A.~H. and A.~S.), the
Polish National Science Centre (NCN) Grant 2023/51/B/ST9/02798
(A.~S. and S.~T.) and the Deutsche For\-schungs\-gemeinschaft (DFG)
Grant No. SE 1836/6-1 (A. S).  S.~T. is a member of the IMPRS for
``Quantum Dynamics and Control'' at the Max Planck Institute for the
Physics of Complex Systems, Dresden, Germany, and acknowledges its
partial support.

\bibliographystyle{Frontiers-Vancouver}

\end{document}